\documentclass[useAMS,usenatbib]{mn2e}
\usepackage[normalem]{ulem}

\usepackage{tabularx,ragged2e,booktabs,caption}
\usepackage{amsmath}
\usepackage{amssymb}
\usepackage{threeparttable}
\usepackage{subfigure}
\usepackage{cases}
\usepackage{color}
\usepackage{graphicx}
\usepackage{hyperref}
\usepackage[figure]{hypcap}
\usepackage{tikz}

\hypersetup{colorlinks, citecolor=blue, pdfduplex=DuplexFlipLongEdge}
\hypersetup{pdftitle=ULX paper, pdfauthor=Ramesh Narayan, pdfsubject=Astrophysics}

\title[Spectra of ULX Sources]
  {Spectra of Black Hole Accretion Models of Ultra-Luminous X-ray Sources}
\author[R. Narayan et al.]
  {Ramesh Narayan$^1$\thanks{\hbox{E-mail: rnarayan@cfa.harvard.edu; asadowsk@mit.edu}; Roberto.Soria@curtin.edu.au
},
 Aleksander S\c{a}dowski$^2$\footnotemark[1], Roberto Soria$^{3,4}$\footnotemark[1]
\\
  $^1$Harvard-Smithsonian Center for Astrophysics, 60 Garden Street, Cambridge, MA 02138, USA\\
  $^2$MIT Kavli Instiatute for Astrophysics and Space Research, 77 Massachusetts Ave, Cambridge, MA 02139, USA\\
  $^3$ International Centre for Radio Astronomy Research, Curtin University, GPO Box U1987, Perth, WA 6845, Australia\\
  $^4$ Sydney Institute for Astronomy, School of Physics A28, The University of Sydney, NSW 2006, Australia}

\begin{document}

\maketitle

%====================================================================================================

\begin{abstract}

We present general relativistic radiation MHD simulations of
super-Eddington accretion on a $10M_\odot$ black hole. We consider a
range of mass accretion rates, black hole spins, and magnetic field
configurations. We compute the spectra and images of the models as a
function of viewing angle, and compare them with the observed
properties of ultraluminous X-ray sources (ULXs). The models easily
produce apparent luminosities in excess of $10^{40}~{\rm erg\,s^{-1}}$
for pole-on observers. However, the angle-integrated radiative
luminosities rarely exceed $2.5\times10^{39}~{\rm erg\,s^{-1}}$ even
for mass accretion rates of tens of Eddington. The systems are thus
radiatively inefficient, though they are energetically efficient when
the energy output in winds and jets is also counted. The simulated
models reproduce the main empirical types of spectra --- disk-like,
supersoft, soft, hard --- observed in ULXs.  The magnetic field
configuration, whether MAD (``magnetically arrested disk") or SANE
(``standard and normal evolution"), has a strong effect on the
results.  In SANE models, the X-ray spectral hardness is almost
independent of accretion rate, but decreases steeply with increasing
inclination.  MAD models with non-spinning black holes produce
significantly softer spectra at higher values of $\dot{M}$, even at
low inclinations.  MAD models with rapidly spinning black holes are
quite different from all other models. They are radiatively efficient
(efficiency factor $\sim10-20\%$), super-efficient when the mechanical
energy output is also included (70\%), and produce hard blazar-like
spectra.  In all models, the emission shows strong geometrical
beaming, which disagrees with the more isotropic illumination favoured
by observations of ULX bubbles.

\end{abstract}

%====================================================================================================
%====================================================================================================

\section{Introduction}\label{sec:intro}

Ultraluminous X-ray sources (ULXs) are a class of highly luminous,
compact, non-nuclear X-ray sources in nearby spiral galaxies, whose
luminosities exceed the Eddington luminosity limit of a neutron star,
or even that of a $10M_\odot$ black hole
\citep{fabbiano89,makishima00,swartz04}. The nature of these
mysterious sources is still not understood.

Because of their very large apparent luminosities ---
few$\times10^{39}~{\rm erg\,s^{-1}}$ to above $10^{40}~{\rm
  erg\,s^{-1}}$ in a few cases --- it was suggested that ULXs might be
intermediate mass black holes (BHs) \citep{miller04}. While one or two
ULXs may well be intermediate mass BHs (e.g., HLX-1:
\citealt{farrell09,godet09,davis10}), the more recent consensus (see
\citealt{bachetti16,feng11} for reviews, and
\citealt{king01,begelman06,poutanen07} for theoretical arguments) is
that the vast majority of ULXs are stellar-mass ($\sim10M_\odot$) BHs,
accreting above their Eddington limit. But not all ULXs are BHs: three
objects show coherent pulsations and are thus neutron stars
\citep{bachetti14,israel17a,israel17b,furst17,furst16}. It is unclear
what fraction of ULXs belong to this class \citep{king16}.

The identification of accreting neutron stars within the ULX
population implies that accreting systems can certainly have highly
super-Eddington apparent luminosities. Two alternative physical
scenarios might explain this fact.  One possibility is that the photon
emission of super-critical neutron stars and black holes is strongly
collimated along the polar axis, and appears highly super-Eddington
only for observers located in that direction; a beaming factor scaling
as $\left(\dot{M}/\dot{M}_{\rm Edd}\right)^2$ was proposed by
\cite{king09}. Another scenario \citep{israel17b}, specific to neutron
stars, is that the classical Eddington limit is not a barrier to
accretion onto a highly magnetized neutron star because the electron
scattering cross section (and therefore the effect of radiation
pressure) is reduced for photon energies in the X-ray band, in the
presence of a magnetic field $B \ga 10^{12}$ G \citep{herold79}.

Regardless of details, an unavoidable consequence of super-critical
accretion is that the inflow cannot settle into a standard,
radiatively efficient, optically thick, geometrically thin accretion
disk \citep{shakura73}. A generalization of the thin disk, the ``slim
disk'' model \citep{abramowicz88} has been widely applied to the
regime of super-Eddington accretion, including ULXs (e.g.,
\citealt{watarai01}). It is a useful first approximation in the study
of such objects. However, these idealized analytic models are not
appropriate for detailed comparison with observations, as it is
generally believed that super-Eddington flows will have massive
radiatively-driven outflows. Such outflows are intrinsically
two-dimensional and cannot be understood within a height-integrated 1D
accretion framework. The winds will cause anisotropic emission, with
geometric collimation of the radiation along the polar axis. The gas
in the wind will also scatter the radiation from the disk. Predicting
the spectral appearance and apparent luminosity of a super-critical
accretor, as seen by distant observers, is therefore a challenging
problem that requires numerical simulations.

Pioneering work on simulating super-Eddington accretion disks was done
by Ohsuga and collaborators, who developed radiation-hydrodynamic
\citep{ohsuga05} and radiation-MHD codes \citep{ohsuga11}. Using an
axisymmetric two-dimensional model with a pseudo-Newtonian potential
\citep{paczynsky80}, \cite{kawashima12} calculated the apparent
luminosity and spectral appearance of super-Eddington sources for
different viewing angles; their calculations included bulk and thermal
Compton upscattering of seed disk photons in a hot (shocked) inner
region, and Compton downscattering and absorption through a dense
outflow. The authors obtained good agreement between their model
predictions and spectra observed in some ULXs.

During the last few years, fully general relativistic radiation MHD
codes have been developed by a number of groups
\citep{sadowski13,sadowski14,mckinney14,fragile14,takahashi15,takahashi16}.
In this paper, we use one of these state-of-the-art codes,
\texttt{KORAL} \citep{sadowski13,sadowski14}, to explore the
super-Eddington accreting stellar-mass BH model of ULXs. We present a
number of general relativistic radiation MHD simulations of accreting
BHs, corresponding to a range of super-Eddington mass accretion rates,
BH spins and magnetic field strengths. We compute spectra and images
corresponding to these simulations using a radiative-transfer and
ray-tracing code \texttt{HEROIC} \citep{narayan16,zhu15}. We then
discuss to what extent the numerical accretion models reproduce the
observed spectra of ULXs. 
%and whether the opening angle of the polar funnel is a strong
%function of accretion rate, as proposed by \cite{king09}.

While ULXs are of great interest in and of themselves, they are also
convenient prototypes of super-Eddington accretion flows in other more
distant objects. It is believed that many tidal disruption events
(TDEs) go through a super-Eddington phase at early times
\citep{alexander16,socrates12,lodato11,rees88,zauderer11}. A sub-class
of active galactic nuclei (AGN) in the local universe, known as Narrow
Line Seyfert 1 galaxies, are likely close to the Eddington limit and
in some cases probably super-Eddington
\citep{jin16,castello-mor16,zubovas13,kawataku11,collin04}.  Finally,
the rapid early growth of supermassive BHs, as evidenced by the
presence of very massive BHs at high redshifts
\citep{wu15,zuo15,mortlock11}, and also from direct measurements of
the luminosity from the most powerful quasars
\citep{wang15,page14,kelly13}, might indicate that these BHs grew via
a super-Eddington phase in the early universe
\citep{lupi16,volonteri15,volonteri05,madau14,king03}.  Progress in
these fields will be possible only when we develop tools for studying
super-Eddington accretion and understand the nature of such flows.

Longstanding questions on the nature of super-Eddington accretion
include: (i) How viable is super-Eddington accretion in the first
place? (We now know that it is certainly viable because some ULXs have
turned out to be accreting neutrons stars.)  (ii) What is the geometry
of the accretion flow? How does it impact observations as a function
of inclination angle?  (iii) How luminous are super-Eddington systems?
Are they radiatively efficient?  (iv) How much mechanical energy do
super-Eddington disks produce in outflows? What role do the outflows
play in feedback?  (v) How often do super-Eddington disks produce
relativistic jets? How do these jets compare with blazar jets?

ULXs have several advantages for exploring these basic questions.
Even though we do not have a precise BH mass measurement for any
individual ULX, it is reasonable to assume that the mass of a typical
ULX (the non-neutron star variety) is not very different from
$\sim10M_\odot$ (by no more than a factor of 2-3). This eliminates one
large source of uncertainty. ULXs exhibit at least four different
spectral states, which indicates that the complex physics of
super-Eddington accretion is well-represented by this population.  In
a few sources, transitions between spectral states have been observed,
which is likely to be helpful for understanding the origin of the
different states. ULXs have bubbles of ionized gas surrounding them,
which provide information on the net angle-integrated outflow of
radiation and mechanical energy from the accreting BH. This gives
independent constraints on the isotropic energy output of the system,
as distinct from any geometrically focused radiation that may be
received directly from the accretion disk. The present study
represents a first effort at understanding these and other
observations of ULXs.

In \S\ref{sec:numerical}, we describe the numerical methods used in
this work, specifically, the general relativistic radiation MHD
(GRRMHD) code \texttt{KORAL} and the radiation post-processing code
\texttt{HEROIC}. In \S\ref{sec:fiducial}, we discuss results for our
fiducial model, which consists of a $10M_\odot$ non-spinning BH,
accreting at 10 times the Eddington mass accretion rate. In
\S\ref{sec:parameters}, we carry out a parameter study, where we
compare models with different mass accretion rates, BH spins, and
magnetic field strengths.  In \S\ref{sec:ULX_comparison}, we compare
the simulation results and computed spectra with observations of
ULXs. The comparison is promising, but there are also clear
discrepancies.  Finally, in \S\ref{sec:discussion}, we conclude with a
summary and discussion.

\section{Numerical Methods}
\label{sec:numerical}

The computations discussed in this paper are done in three stages, as
described in the following subsections. First, we run a GRRMHD
simulation of the accretion flow for the chosen model
parameters. Next, we transfer the simulation output to a second grid
and extrapolate the data to large radii, where the GRRMHD data have
not converged. Finally, we solve for the radiation field on the second
grid using a post-processing code

\subsection{GRRMHD simulations with \texttt{KORAL}}\label{sec:grrmhd}

The simulations were done using the GRRMHD code \texttt{KORAL}
\citep{sadowski13,sadowski14}, which evolves gas, magnetic field and
radiation in a fixed gravitational metric. In the present work, we use
the Kerr metric in Kerr-Schild coordinates. The magnetic field is
evolved assuming ideal MHD (no resistivity) and the radiation is
described by means of frequency-integrated angular moments, with the
moment expansion closed via the M1 closure method
\citep{levermore84}. A radiative viscosity term is included in order
to mitigate some of the limitations of the M1 scheme
\citep{sadowski15a}.

The radiative processes included in the present simulations are
free-free emission and absorption, and Compton scattering, where the
latter is handled via a photon-conserving scheme
\citep{sadowski15b}. \texttt{KORAL} and its sister code HARMRAD
\citep{mckinney14} are capable of modeling additional radiation
processes such as thermal synchrotron \citet{sadowski17} and double
Compton \citet{mckinney17}), but these were not included in the
present work.

Table \ref{tab:models} lists the key parameters of the 13 simulations
discussed in this paper.  All models assume a BH mass
$M=10M_\odot$. Of the 13 simulations, 6 have been done in 3D, where
the magnetorotational instability (MRI, \citealt{balbus91,balbus98})
is well-resolved and develops robustly, while 7 are in 2D. It is
well-known that the MRI cannot be sustained in 2D, so 2D MHD
simulations cannot achieve steady state accretion. To overcome this
problem, we employ the mean-field magnetic dynamo prescription
described in \citet{sadowski15a}, which permits us to run 2D
simulations for arbitrarily long times. Previous tests have shown that
such 2D simulations agree well with their 3D counterparts, at least in
their time-averaged properties \citep{sadowski15c}, and are an
economical way of running simulations.

All simulations were initialized with an equilibrium torus of weakly
magnetized gas orbiting the BH.  The torus parameters are chosen to
correspond closely to the initial setup of earlier simulations
described in \citet{sadowski16a}. The initial gas density in the torus
is adjusted to obtain the desired mass accretion rate.  The topology
of the initial seed magnetic field is also adjusted, depending on the
requirements. In the case of MAD (``Magnetically Arrested Disk'',
\citealt{igumenshchev03,narayan03,tchekhovskoy11}) models, we
initialize the simulation with a single large-scale loop, while for
SANE (``Standard and Normal Evolution'', \citealt{narayan12}) models,
we use multiple loops of alternating polarity.

The 2D simulations were run with a resolution of 320x320 cells in
$r$-$\theta$, and the 3D runs with a resolution of 320x320x32, with 32
cells in azimuth spanning a $\pi/2$ wedge, with periodic boundary
conditions. The adopted grid is logarithmic in $r$ and slightly biased
towards the equatorial plane in $\theta$. Every simulation is run
until a final time of $25,000\,GM/c^3$, which in most cases gives a
well converged solution extending up to $r\sim 30GM/c^2$ at the
equatorial plane, and much farther out at higher latitude.

\begin{table}
\begin{center}
\caption{List of simulated models}
\label{tab:models}
\begin{tabular}{lcccc}
\hline
\hline
Model & $\dot{M}/\dot{M}_{\rm Edd}$ &
$a_*$ & Field Strength & 2D/3D \\
\hline
\hline
\texttt{\bf r010\_3d} & {\bf 10} & {\bf 0} & {\bf SANE} & \bf {3D} \\
\hline
\texttt{r010\_2d} & 10 & 0 & SANE & 2D \\
\texttt{r012\_3d} & 1.2 & 0 & SANE & 3D \\
\texttt{r030\_2d} & 7.0 & 0 & SANE & 2D \\
\texttt{r031\_2d} & 17 & 0 & SANE & 2D \\
\hline
\texttt{r011\_2d} & 12 & 0.9 & SANE & 2D \\
\texttt{r032\_2d} & 6.2 & 0.9 & SANE & 2D \\
\texttt{r033\_2d} & 10 & 0.9 & SANE & 2D \\
\texttt{r034\_2d} & 26 & 0.9 & SANE & 2D \\
\hline
\texttt{r013\_3d} & 23 & 0 & MAD & 3D \\
\texttt{r023\_3d} & 1.3 & 0 & MAD & 3D \\
\hline
\texttt{r014\_3d} & 36 & 0.9 & MAD & 3D \\
\texttt{r015\_3d} & 6.8 & 0.9 & MAD & 3D \\
\hline
\hline
\end{tabular}
\end{center}
\end{table}

The 3D model r010\_3d, shown in bold in the first line of
Table~\ref{tab:models}, is our {\it fiducial} model. It considers a
non-spinning BH, $a_* \equiv a/M = 0$, and has a mass accretion rate
$\dot{M}=10\dot{M}_{\rm Edd}$. In this paper, we define $\dot{M}_{\rm
  Edd}$ in terms of the Eddington luminosity, $L_{\rm Edd} = 1.25
\times 10^{38} (M/M_\odot)\,{\rm erg\,s^{-1}}$, for the given BH mass
$M$,
\begin{equation}
\dot{M}_{\rm Edd} = \frac{L_{\rm Edd}}{\eta_{\rm NT} c^2},
\label{eq:edd}
\end{equation}
where $\eta_{\rm NT}$ is the radiative efficiency of the
\citet{novikov_thorne73} general relativistic thin accretion disk
model. For $a_*=0$, $\eta_{\rm NT}=0.05719$, while for $a_*=0.9$
(below), $\eta_{\rm NT}=0.1558$. The fiducial model is initialized
with a weak poloidal magnetic field in multiple loops such that, even
after the disk has reached steady state for a considerable period of
time, the poloidal field strength at the BH horizon is still at the
SANE level.

Model r010\_2d is identical to the fiducial model, but it is
run in 2D. This model is used to verify that results in 2D are close
to those obtained in 3D. 

Models r012\_3d, r030\_2d and r031\_2d are similar to the previous two
models in that they have $a_*=0$ and a SANE magnetic field, but their
mass accretion rates are different, as indicated in
Table~\ref{tab:models}.

Models r011\_2d, r032\_2d, r033\_2d and r034\_2d correspond to
spinning BHs, with $a_*=0.9$. These models cover a range of values of
$\dot{M}$, and all have SANE magnetic fields.

The final 4 models in Table \ref{tab:models} were initialized with a
single poloidal loop of magnetic field and therefore ended up with
strong poloidal magnetic fields, corresponding to the MAD
limit. Models r013\_3d and r023\_3d have a non-spinning BH, $a_*=0$,
while models r014\_3d and r015\_3d have a spinning BH, $a_*=0.9$. All
MAD models have significant non-axisymmetric structure
\citep{igumenshchev03,tchekhovskoy11,mckinney12}, and they have to be
run in 3D.

\subsection{Radial extrapolation of simulated models}\label{sec:disk_extension}

Two steps are needed before the \texttt{KORAL} simulation output can
be post-processed by the radiation solver described in
\S\ref{sec:heroic}.

First, time-averaged and azimuth-averaged (in the case of 3D
simulations) data are interpolated on to the grid that will be used
during the post-processing stage. We use 81 cells distributed
uniformly in $\theta$ and 50 cells per decade distributed uniformly in
$\log r$. This gives roughly square cells in $r\theta$, which
minimizes the effects of ray defects \citep{zhu15}. The radial grid
extends from an inner radius just outside the BH horizon to an outer
radius $r_{\rm out} = 10^5GM/c^2$.

The \texttt{KORAL} data are averaged over the chosen time duration,
which is the last $5,000\,GM/c^3$ of each simulation, and over the
full azimuth range of $\pi/2$ in the case of 3D simulations. The data
re then transferred to the new grid by simple linear interpolation. In
the case of the viscous heating rate $q^+$, we ignore the
\texttt{KORAL} values in the four cells in $\theta$ closest to the
poles, where boundary conditions make the results unreliable, and
instead extrapolate from larger $\theta$.

The second step is to cut out regions of the \texttt{KORAL} simulation
that are outside the converged region of the simulation and to
extrapolate the \texttt{KORAL} data to these cells. For each cell in
the interpolated grid, we compute the poloidal velocity, $v_{\rm pol}
= (v_r^2+v_\theta^2)^{1/2}$, and compute a flow time $t_{\rm flow} =
r/v_{\rm pol}$. We then compare $t_{\rm flow}$ to a characteristic
simulation duration $t_{\rm sim}$ of the \texttt{KORAL}
simulation. For the latter, we use either $t_{\rm sim} =
5,000\,GM/c^3$, the duration over which the simulation output is
time-averaged, or $t_{\rm sim} = 12,500\,GM/c^3$, half the total
duration of the simulation (they give similar results).
%\sout{(see Table ?)}. 

If $t_{\rm flow} < t_{\rm sim}$, we consider the fluid in the cell in
the \texttt{KORAL} simulation to have reached steady state. After
identifying all the cells in the grid that are in steady state
requirement, for each $\theta$, we call the outermost radius that
satisfies this condition as the limiting equilibrium radius $r_{\rm
  eq}(\theta)$. Cells with $r>r_{\rm eq}(\theta)$ have $t_{\rm flow} >
t_{\rm sim}$ and are less likely to have achieved steady state.

\begin{figure*}
\includegraphics[width=1.13\columnwidth]{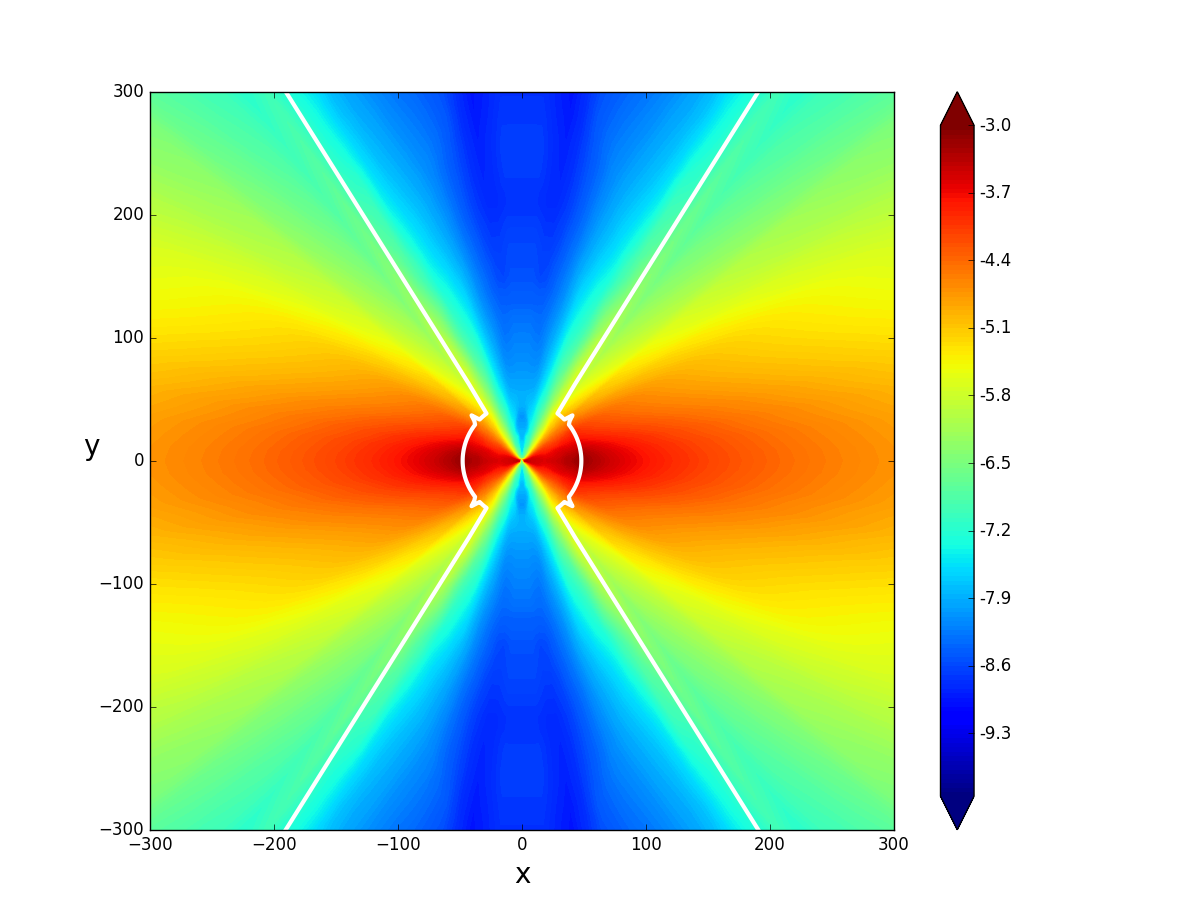}\hspace{-1.5cm}
\includegraphics[width=1.13\columnwidth]{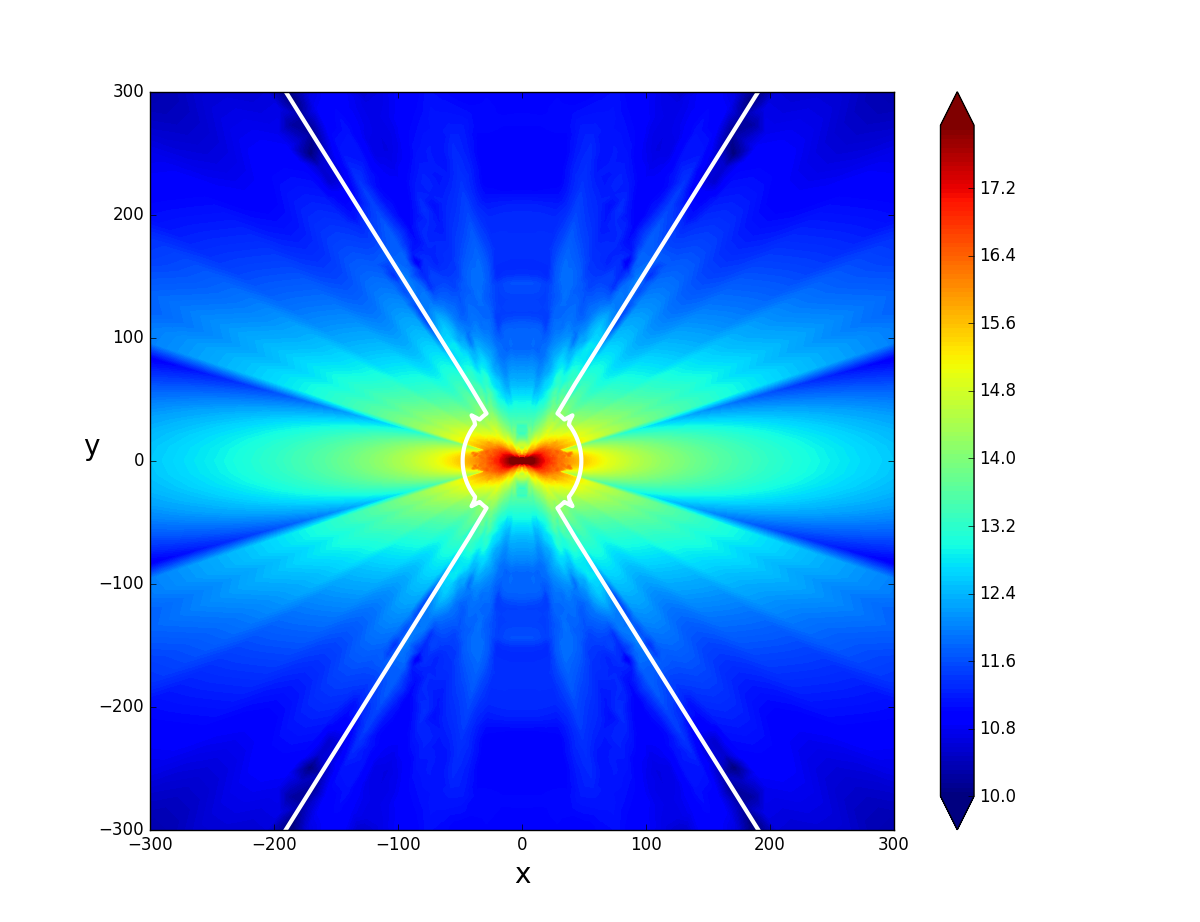} \\
\caption{The two panels show the distribution of the logarithm of the
  density $\log\rho$ (left) and logarithm of the viscous heating rate
  $\log q^+$ (right) in the poloidal plane of the fiducial model
  r010\_3d. The BH is at the center and coordinates are expressed in
  mass units ($GM/c^2$). The disk equatorial plane is oriented
  horizontally and the two polar funnels are oriented vertically. The
  regions inside the white contour are in steady state in the
  \texttt{KORAL} simulation. Outside the white contours, the
  \texttt{KORAL} data are extrapolated, as explained in
  \S\ref{sec:disk_extension}.}
\label{fig:extrapolate}
\end{figure*}

The white contour in Figure~\ref{fig:extrapolate} shows the boundary
of the steady-state region, $r_{\rm eq}(\theta)$, for the fiducial
model. In the equatorial regions, steady state is achieved out to
radii $\sim 30GM/c^2$, but we arbitrarily set $r_{\rm
  eq}(\theta)=50GM/c^2$. The \texttt{KORAL} simulation data are
somewhat less reliable in the radius range $r/(GM/c^2) \sim 30-50$,
but we feel it is better to use the simulation data here rather than
purely extrapolated data since there is non-negligible viscous
dissipation at these radii. In addition, there is non-negligible
radial advection of radiation in a couple of the models, and it is
hard to model advection correctly in extrapolated data. In the polar
regions, because of the large velocity of outflowing gas, steady state
is achieved to much larger radii. In fact, for angles within about
$30^\circ$ of the poles, the flow is in steady state out to the edge
of the \texttt{KORAL} simulation at $r=10^3GM/c^2$. To avoid edge
effects, we ignore the last 5 radial cells in the \texttt{KORAL}
output, so this limits $r_{\rm eq}(\theta)$ to around $900GM/c^2$ in
the polar regions.

For cells with $r > r_{\rm eq}(\theta)$, we extrapolate from the
\texttt{KORAL} values at $r = r_{\rm eq}(\theta)$, using an appropriate
scaling as a function of $r$.  In all our simulations, the accretion
flow has two distinct regions: (i) An inflow region which is
restricted to a range of angles around the equator, and (ii) an
outflow region which consists of higher latitudes, extending up to the
poles. For each quantity that we extrapolate, we first identify which
of these two zones is more important to model correctly. We then
choose a radial scaling appropriate for that zone, but apply it to the
entire extrapolated volume. Although the scaling may be inconsistent
for the other zone, it generally does not matter.  With this idea in
mind, the scalings we use for the extrapolated region, $r>r_{\rm
  eq}(\theta)$), are as follows:
\begin{eqnarray}
\rho(r,\theta) &=& \rho[r_{\rm eq}(\theta)] \left[\frac{r}{r_{\rm
      eq}(\theta)}\right]^{-2}, \label{eq:rhoscale} \\ 
v_r(r,\theta) &=& v_r[r_{\rm eq}(\theta)]\,, \label{eq:vrscale} \\ 
v_\theta(r,\theta) &=& v_\theta[r_{\rm eq}(\theta)]
\left[\frac{r}{r_{\rm eq}(\theta)}\right]^{-1/2}, \label{eq:vthetascale} \\ 
v_\phi(r,\theta) &=& v_\phi[r_{\rm eq}(\theta)] \left[\frac{r}{r_{\rm
      eq}(\theta)}\right]^{-1/2}, \label{eq:vphiscale} \\ 
T_{\rm gas}\,(r,\theta) &=& T_{\rm gas}\,[r_{\rm eq}(\theta)]
\left[\frac{r}{r_{\rm eq}(\theta)}\right]^{-1}, \label{eq:tgasscale} \\ 
q^+(r,\theta) &=& q^+[r_{\rm eq}(\theta)] \left[\frac{r}{r_{\rm
      eq}(\theta)}\right]^{-4} ~{\rm (poles)}, \label{eq:qplusscale} \\ 
|B|\,(r,\theta) &=& |B|\,[r_{\rm eq}(\theta)] \left[\frac{r}{r_{\rm
      eq}(\theta)}\right]^{-3/2}.  \label{eq:bscale}
\end{eqnarray}

The scalings for $\rho$ and $v_r$ (eqs.~\ref{eq:rhoscale},
\ref{eq:vrscale}) are driven by our desire to model the outflow
density and dynamics correctly. Specifically, we wish the outflowing
gas to coast at a constant radial velocity (which is reasonable
because $r_{\rm eq}$ is usually large enough that we are outside the
acceleration zone of the outflow), and to conserve mass. These
scalings are not accurate for the inflowing equatorial disk, but we
believe that the error we make is unimportant since the gas here has a
very low radial velocity and is optically
thick. Figure~\ref{fig:extrapolate} shows the density distribution we
obtain via this extrapolation technique for the fiducial model.

For $v_\theta$ and $v_\phi$, we use a Keplerian scaling with radius
(eqs.~\ref{eq:vthetascale}, \ref{eq:vphiscale}), as appropriate for
gas orbiting in the disk. This is particularly important for $v_\phi$,
which can produce significant Doppler shifts even at largish radii. In
the outflow region, $v_\theta$ and $v_\phi$ are much smaller than
$v_r$, and it does not matter what scaling we use.

For the gas temperature $T_{\rm gas}$, we use a virial argument to
choose a radial scaling $\propto r^{-1}$ (eq.~\ref{eq:tgasscale}). The
precise choice is not important since, in most of our radiation
modeling, we solve self-consistently for the gas temperature
(\S\ref{sec:heroic}).

We scale the viscous heating rate $q^+~({\rm erg\,cm^{-3}s^{-1}})$ as
$r^{-4}$ (eq.~\ref{eq:qplusscale}). This is demanded by the
requirement that $r^3q^+$ should vary as $r^{-1}$, the fractional
energy released down to radius $r$. In practice, we extrapolate only
at polar angles, where the amount of heating involved is not large. In
the equatorial region, we use thin disk theory to determine the amount
of energy dissipated as a function of radius and angle. Specifically,
we take the dissipation rate per unit area $Q^+$ ($\rm erg\,cm^{-2}$)
of a non-relativistic thin disk of given $M$ and $\dot{M}$
\citep{shakura73},
\begin{equation}
Q^+(r) = \frac{3GM\dot{M}}{4\pi r^3} \left[1-\left(\frac{r}{r_{\rm
      in}}\right)^{-1/2}\right],
\end{equation}
and distribute it with a gaussian distribution in $\theta$ around the
equatorial plane to model the dissipation rate per unit volume $q^+$
($\rm erg\,cm^{-3}$):
\begin{equation}
q^+(r,\theta) = \frac{Q^+(r)}{\sqrt{2\pi}r\theta_s}\,
\exp\left[-\frac{\left(\theta-(\pi/2)\right)^2}{2\theta_s^2}\right] ~{\rm (equator)}.
\label{eq:qplusthin}
\end{equation}
We use $\theta_s=0.1$ for the angular scale height (the exact value is
unimportant since this heating occurs deep inside the optically thick
portion of the disk), and $r_{\rm in}=6GM/c^2$ for the nominal inner
edge of the thin disk model.  Figure~\ref{fig:extrapolate} shows the
resulting distribution of $q^+$.  Note that the bulk of the viscous
energy release occurs inside the steady state region $r < r_{\rm
  eq}(\theta)$, where we use \texttt{KORAL} results, so the energy
release in the extrapolated region, whether we use the polar
extrapolation (\ref{eq:qplusscale}) or the equatorial extrapolation
(\ref{eq:qplusthin}), is quantitatively small.

Finally, the magnetic field strength $|B|$, which is needed for one
test where we include thermal synchrotron emission, is scaled as
$r^{-3/2}$ (eq.~\ref{eq:bscale}). This is to ensure that the magnetic
pressure $B^2/8\pi$ scales the same way as the gas pressure $\rho
T_{\rm gas}$.

We extrapolate all the above quantities out to a radius $r_{\rm out} =
10^5GM/c^2$. This is perhaps farther out than necessary. However, we feel
that there is value in allowing the radiation model to include opacity
and reprocessing effects at large radii. Since our grid is logarithmic
in radius, the extra cost of handling a large range of radius during
the radiation post-processing step is not excessive.

\subsection{Radiation post-processing with \texttt{HEROIC}}
\label{sec:heroic}

Radiation post-processing is done using the multi-dimensional, general
relativistic code \texttt{HEROIC} \citep{narayan16,zhu15}. This code
takes the density, velocity, viscous dissipation rate and other
quantities in the interpolated grid described in
\S\ref{sec:disk_extension}, and solves in detail for the radiation
field in each grid cell. In the present work, we describe the angular
distribution of the radiation field by solving for the intensity on
162 angles distributed uniformly over the sphere in the local fluid
frame of each cell. We use 101 frequencies, distributed uniformly in
$\log\nu$ from $\nu = 10^{14}-10^{24}$\,Hz, to describe the radiation
spectrum of each angular ray in each spatial cell.

A number of enhancements have been made to \texttt{HEROIC} since
publication of the original methods papers \citep{zhu15,narayan16}. In
brief:

\begin{enumerate}

\item
The treatment of bremsstrahlung in the relativistic regime has been
improved. The emissivity at relativistic temperatures now uses the
formulae given in \citet{ny95}; the corresponding spectral
distribution follows the prescriptions in \citet{gould80}.

\item
For temperatures below $10^8$\,K, the code uses an opacity table
corresponding to solar abundances taken from the the CHIANTI database
\citep{dere97,landi13,delzanna15}; the opacity includes both
bound-free and free-free contributions. However, for simplicity, we
assume that the spectral distribution is the same as for free-free
(i.e., we ignore features like atomic edges).

\item
The previous Comptonization routine in HEROIC \citep{narayan16}, which
was based on solving the Kompaneets equation, has been supplemented
with a relativistic module for temperatures above $10^{8.5}$\,K; this
module uses the Comptonization kernel of \citet{jones68}, with the
corrections given in \citet{coppi90}.

\item
Thermal synchrotron emission and absorption are included, using the
approximate formulae given in \citet{ny95} and \citet{mahadevan96};
this feature is used only in one test in this paper.

\item
The code can handle two-temperature plasmas, including the effects of
advection, as required for simulation output from the
recently-developed two-temperature version of \texttt{KORAL}
\citep{sadowski16b}; this improvement is not needed for the present
work. 

\item
Finally, the code now works with both short and long characteristics
(see \citealt{zhu15}).

\end{enumerate}

The radiative post-processing is done using the interpolated and
extrapolated data described in \S\ref{sec:disk_extension}. The
compuations consist of a number of stages, as described below:

\begin{description}

\item
\noindent{\bf Stage I}: First, we keep the gas temperature $T_{\rm
  gas}$ fixed at the values described in \S\ref{sec:disk_extension},
and we iteratively solve for the radiation field (all angles, all
frequencies, all cells), using the radiative transfer equation and the
method of short characteristics \citep{zhu15,narayan16}.

\item
\medskip\noindent{\bf Stage II}: Next, we relax the constraint on the
gas temperature, and solve simultaneously for both the temperature and
the radiation field, again using short characteristics.  In this
stage, we use the viscous heating rate $q+$ as a constraint and apply
the condition of energy balance \citep{narayan16} to solve for the
temperature. This step is necessary because the \texttt{KORAL}
radiation model is fairly crude (just a few frequency-integrated
angular moments), so the \texttt{KORAL} temperatures are not
reliable. The viscous dissipation on the other hand is likely to be
more robust since it ultimately comes from energy conservation, which
\texttt{KORAL} satisfies well.

\item
\medskip\noindent{\bf Stage III}: Next, we take the solution from the
second stage and improve it with around $10$ iterations of long
characteristics, again solving for both the temperature and the
radiation field.

\item
\medskip\noindent{\bf Stage IV}: Finally, we take the output from the
third stage and carry out ray-tracing to calculate the observed
spectrum and/or image for observers located at various orientations
with respect to the disk.

\end{description}

All of the radiation physics and ray-tracing in \texttt{HEROIC} is
done using general relativistic photon geodesics, including ray
deflections, Doppler shifts and gravitational redshift. Even though
the interpolated grid described in \S\ref{sec:disk_extension} is in 2D
($r$-$\theta$), the radiative transfer calculations are done in 3D,
assuming axisymmetry.\footnote{\texttt{HEROIC} can handle 3D data, but
  this was not used in the present work.}

\section{Fiducial Model}\label{sec:fiducial}

\subsection{Comparing \texttt{KORAL} and \texttt{HEROIC}}

We discuss here in some detail the fiducial model, r010\_3d, which has
a BH with $M=10M_\odot$ and $a_*=0$. The mass accretion rate is
$\dot{M} = 10 \dot{M}_{\rm Edd}$, and the magnetic field strength
corresponds to the SANE regime.

\begin{figure*}
\includegraphics[width=1.13\columnwidth]{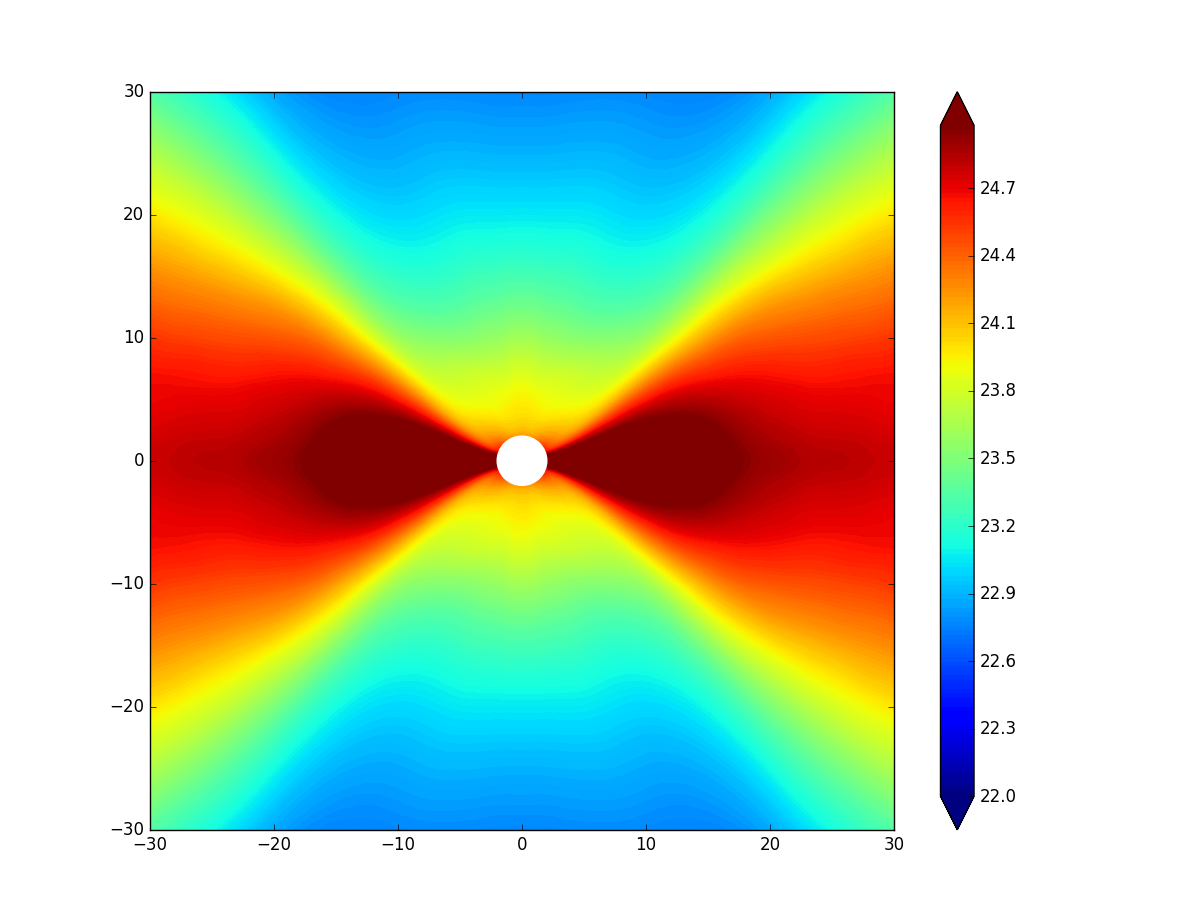}\hspace{-1.5cm}
\includegraphics[width=1.13\columnwidth]{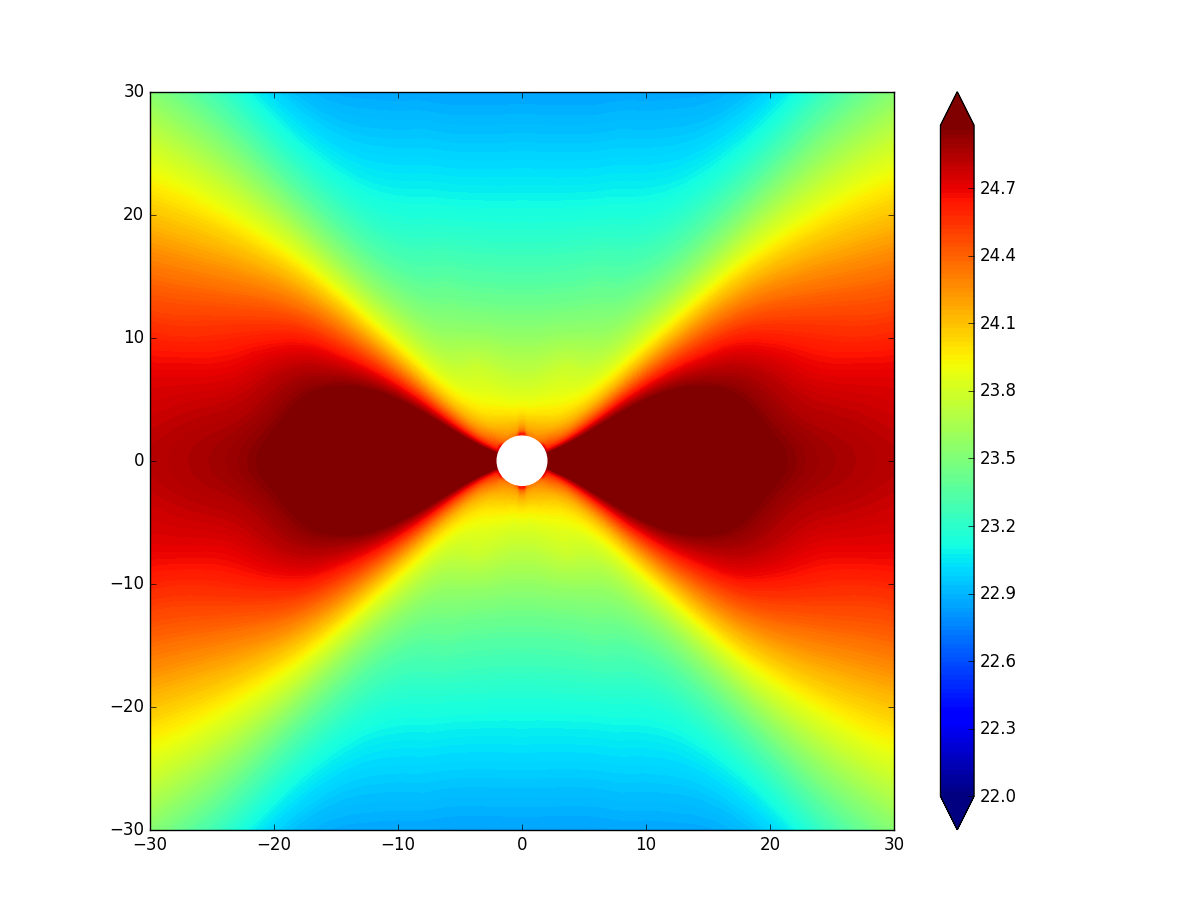} \\
\vspace{-0.25cm}
\includegraphics[width=1.13\columnwidth]{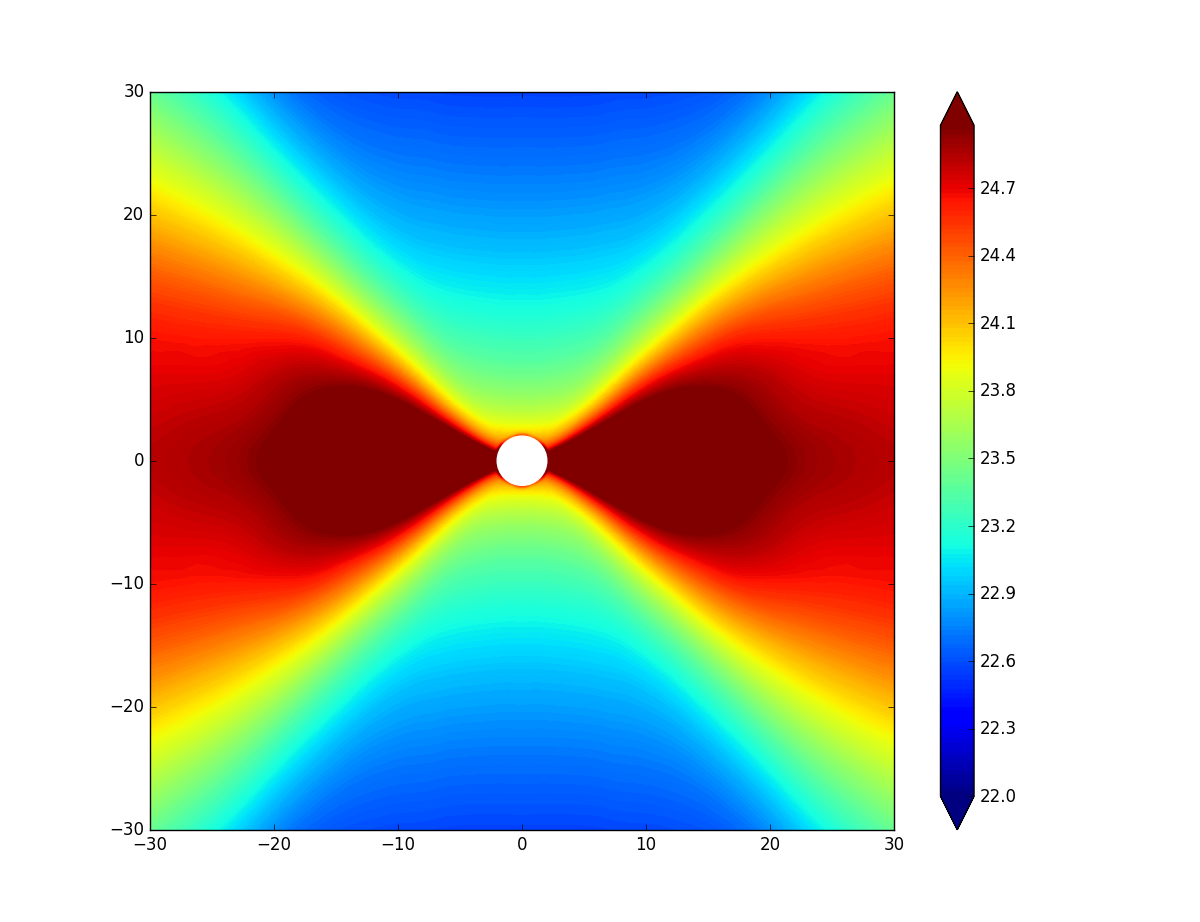}\hspace{-1.5cm}
\includegraphics[width=1.13\columnwidth]{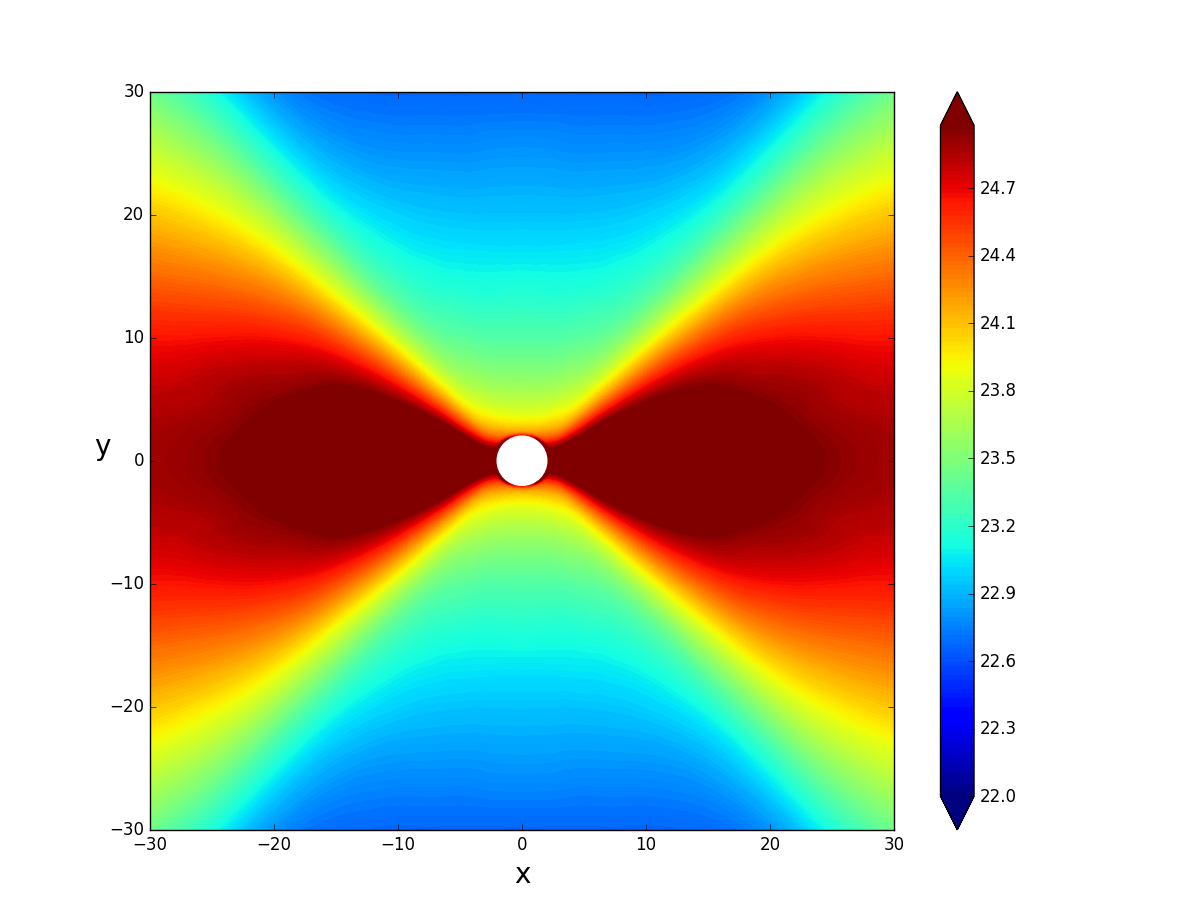}
\vspace{-0.25cm}
\includegraphics[width=1.13\columnwidth]{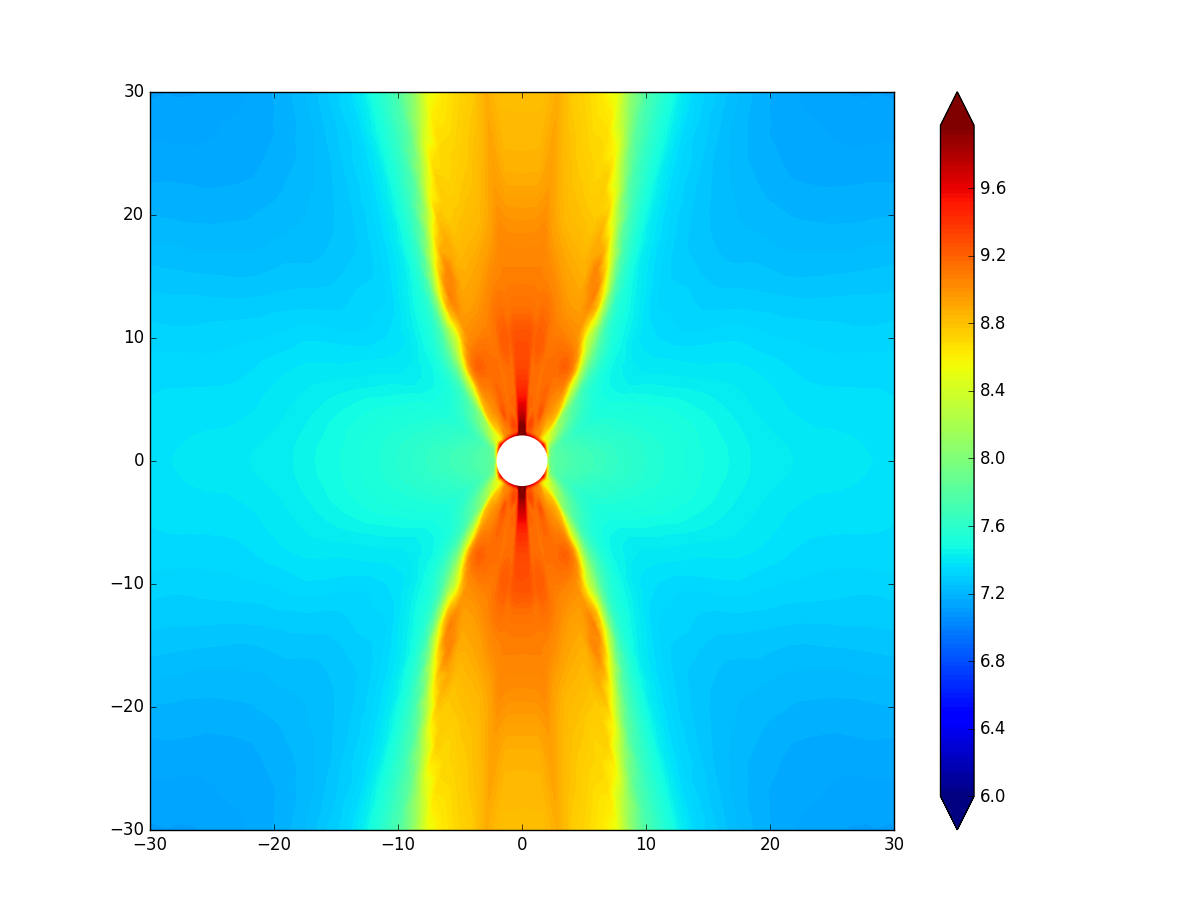}\hspace{-1.5cm}
\includegraphics[width=1.13\columnwidth]{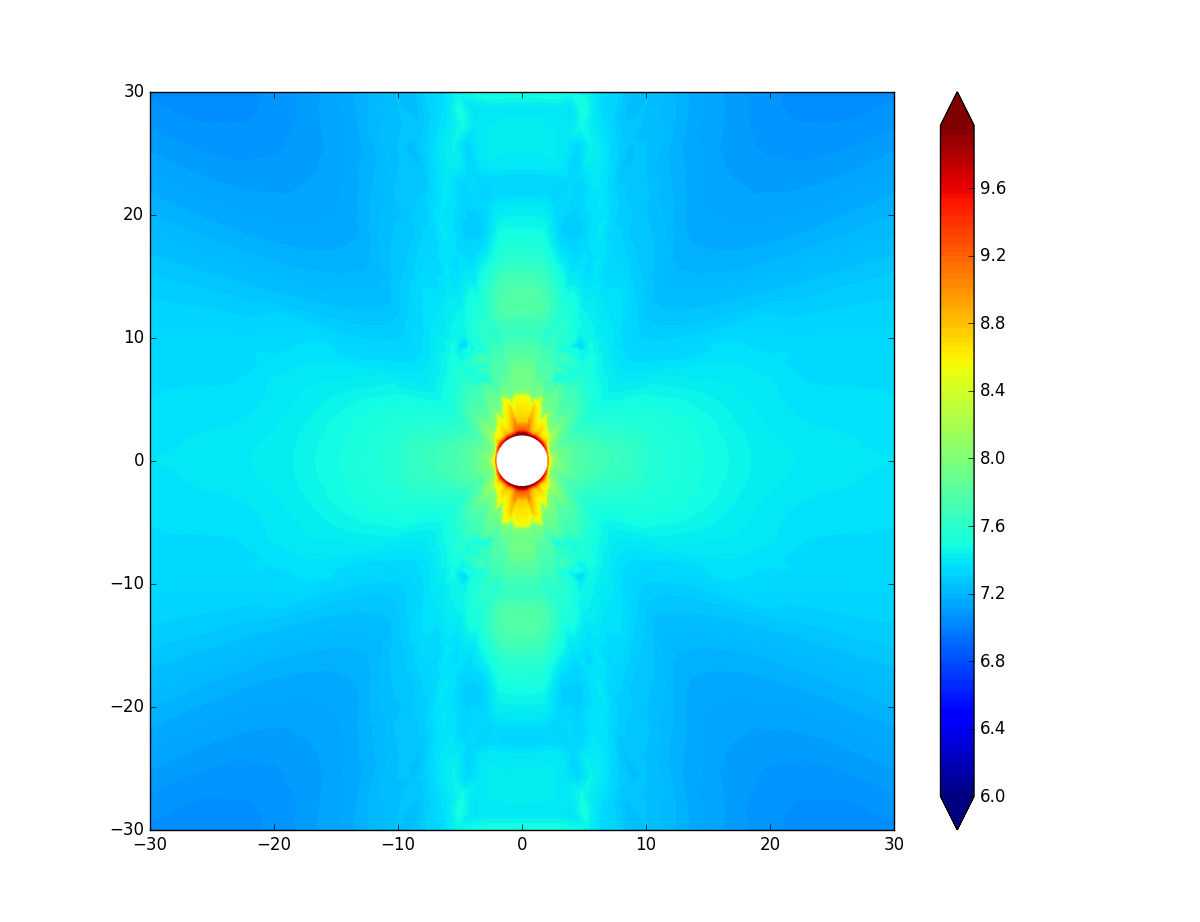} \\
\caption{The top four panels show the mean bolometric radiation
  intensity $\log J$ as a function of position in the poloidal
  plane. Top Left: The distribution of $\log J$ as determined from the
  \texttt{KORAL} simulation of the fiducial model r010\_3d.  Top
  Right: \texttt{HEROIC} solution for $J$ after Stage I, i.e., keeping
  the temperature fixed at the \texttt{KORAL} values and solving only
  for the radiation. Middle Left: \texttt{HEROIC} solution for $J$
  after Stage III, i.e., solving for both the temperature and the
  radiation field. Middle Right: Corresponding \texttt{HEROIC}
  solution after Stage III for the 2D model r010\_2d. Bottom Left:
  Original \texttt{KORAL} solution for the temperature distribution
  ($\log T$) in the 3D model r010\_3d. Bottom Right: \texttt{HEROIC}
  solution for the temperature after Stage III.}
\label{fig:jradint_temp}
\end{figure*}

\texttt{KORAL} works with a very simple description of the radiation,
with only five quantities evolved in each grid cell: the radiation
energy density $\hat{E}$ in the fluid frame, the radiation three-flux
vector $\vec{F}$ in the fluid frame, and the photon number density
$n_r$ in the radiation rest frame \citep{sadowski15b}. From the
time-averaged $\hat{E}$, we obtain the mean bolometric radiation
intensity $J$ in the fluid frame:
\begin{equation}
{\rm \texttt{KORAL}}:\quad J = \frac{c\hat{E}}{4\pi}.
\end{equation}
\texttt{HEROIC} computes the radiation field in detail, solving for
the intensity $I_\nu(\Omega)$ in each cell over 162 ray directions
$\Omega$ and 101 frequencies $\nu$. From this we calculate $J$ by
integrating over frequency and averaging over direction:
\begin{equation}
{\rm \texttt{HEROIC}}:\quad J = \frac{1}{4\pi} \int\int I_\nu(\Omega) \,d\nu\,d\Omega.
\end{equation}

Figure \ref{fig:jradint_temp} shows the radiation and temperature
solutions in the inner region of the flow ($r < 30GM/c^2$) for the
fiducial model as obtained with \texttt{KORAL} and \texttt{HEROIC}.
The \texttt{KORAL} solution for $J$ (Top Left panel) shows an obvious
thick disk plus a wide funnel, as expected for a super-Eddington
accretion flow. The radiation intensity is large inside the optically
thick disk, and is much less in the funnel.  The radiation field shows
some inhomogeneous structure, especially close to the poles.  This is
an artifact introduced by the M1 closure scheme in \texttt{KORAL} (see
the discussion of the ``radiation shock'' effect in
\citealt{sadowski15a}). Although \texttt{KORAL} includes a radiation
viscosity term to mitigate this artifact, it is unable to eliminate it
altogether.

The Top Right panel in Figure~\ref{fig:jradint_temp} shows the
\texttt{HEROIC} solution for $J$ at the end of Stage I, i.e., using
the temperature structure obtained in \texttt{KORAL}, but solving for
the full angular and frequency structure of the radiation field.
\texttt{HEROIC} eliminates some of the inhomogeneities in the polar
radiation field. However, the \texttt{HEROIC} solution ends up with
quite a bit more radiation in the funnel compared to the
\texttt{KORAL} solution, especially at angles around
$30^\circ-40^\circ$ from the axis. In fact, this model produces
significantly more radiation than the viscous dissipation requires and
is thus much too luminosity. The reason for this can be understood as
follows.  Because the \texttt{KORAL} solution had a mild deficit of
radiation near the poles, Compton-cooling was less efficient.
Therefore, \texttt{KORAL} introduced a fairly large gas temperature in
order to produce the necessary Compton-cooling to balance the viscous
heating. \texttt{HEROIC} does not have a deficit of radiation at the
poles. If we insist on using the same temperature as \texttt{KORAL}
obtained, as is done in Stage I, then the resulting Compton-cooling is
too strong and the funnel produces too much luminosity.

The above discrepancy is fixed when we solve self-consistently for the
gas temperature with \texttt{HEROIC} so as to match the viscous
heating rate.  The Middle Left panel in Figure~\ref{fig:jradint_temp}
shows the result we obtain after Stage III. Notice that the radiation
field is smooth in the funnel, with no trace of any inhomogeniety.  At
the same time, the overall radiation intensity in the funnel is fairly
well-matched to the \texttt{KORAL} result (Top Left), and much less
that of the \texttt{HEROIC} Stage I result (Top Right). The
corresponding change in the temperature in the funnel is fairly large,
as can be seen by comparing the \texttt{KORAL} temperatures (Bottom
Left panel) and the self-consistent \texttt{HEROIC} temperatures
(Bottom Right).  The differences are primarily in the funnel, whereas
the temperature in the disk interior is hardly changed.

The above comparison shows that, in radiatively efficient regions of
the accretion flow, it is preferable to solve for the temperature
self-consistently with \texttt{HEROIC}. Because \texttt{KORAL} uses a
moment method and M1 closure, it does not include enough degrees of
freedom in its description of the radiation field to obtain accurate
results. Presumably, a more ambitious radiation scheme, such as a
general relativistic version of the variable Eddington tensor (VET)
method described in \citet{jiang12}, will perform better. Meanwhile,
working with \texttt{KORAL}, we find that it is necessary to
post-process with \texttt{HEROIC} up to Stage III, or at least Stage
II, if we wish to have a consistent description of the radiation field
in the funnel. The optically thick and advective regions of the flow
do not require such care.

\begin{figure*}
\includegraphics[width=1.13\columnwidth]{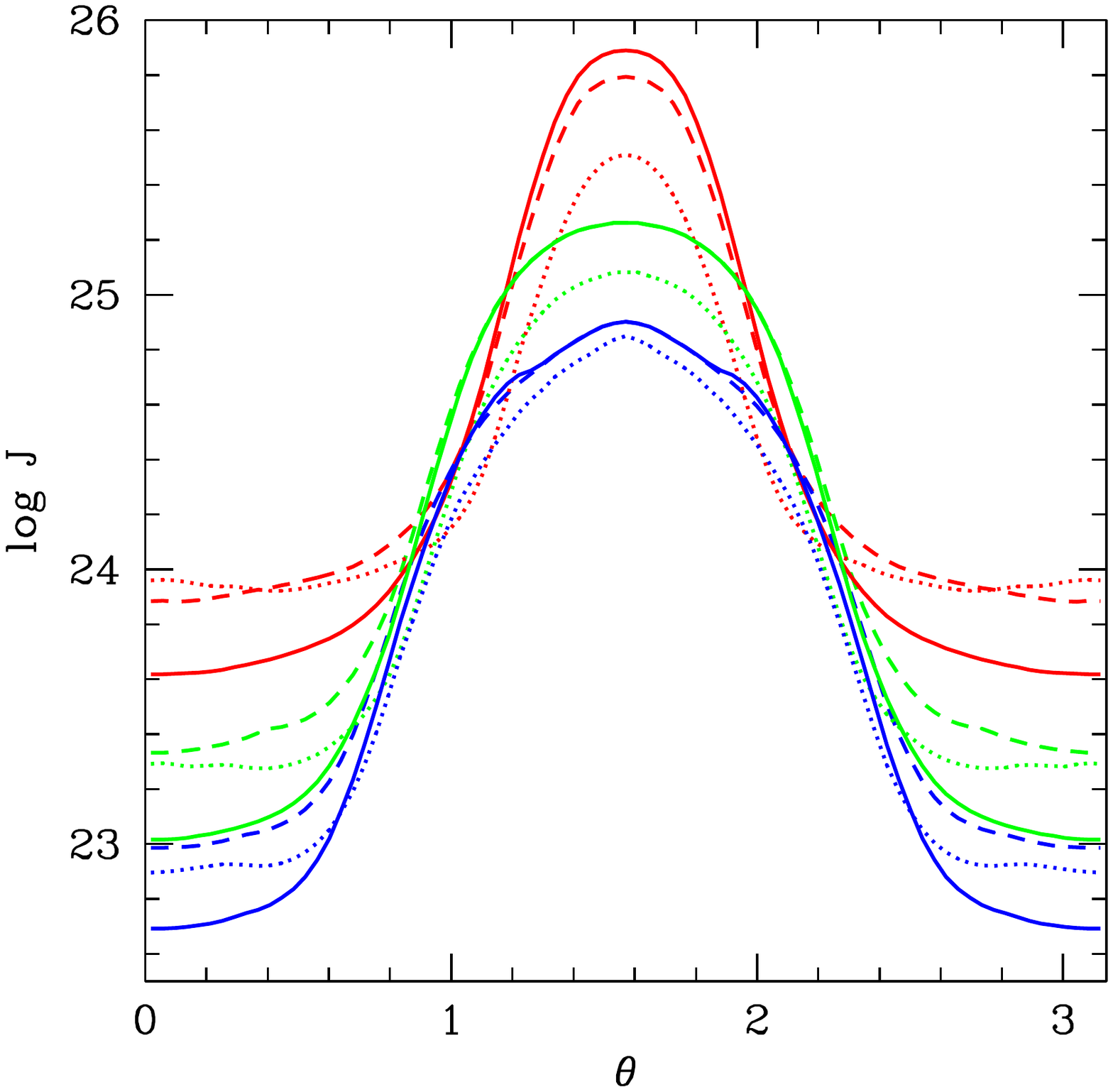}\hspace{-1.5cm}
\includegraphics[width=1.13\columnwidth]{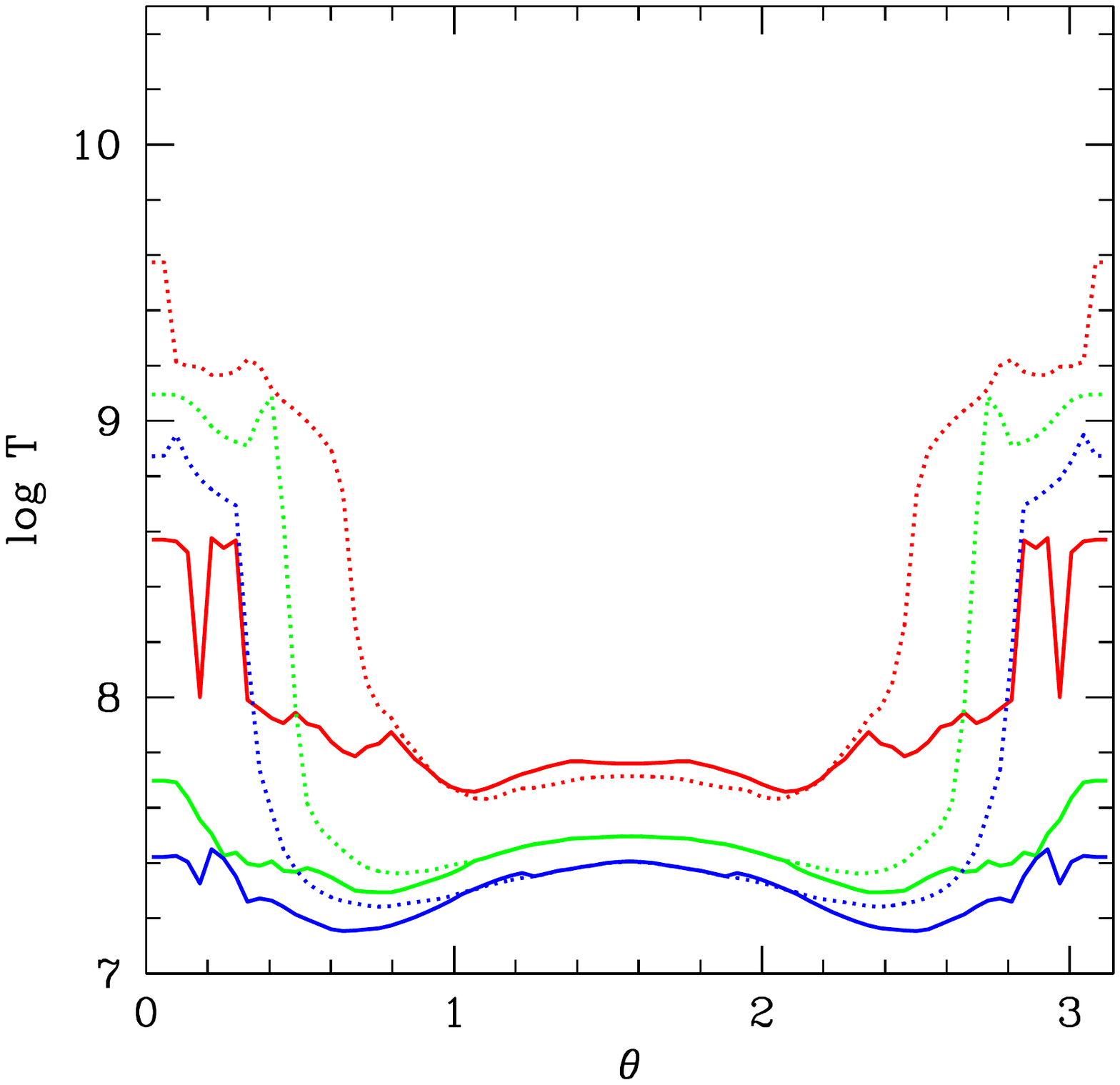} \\
\vspace{-4cm}
\includegraphics[width=1.13\columnwidth]{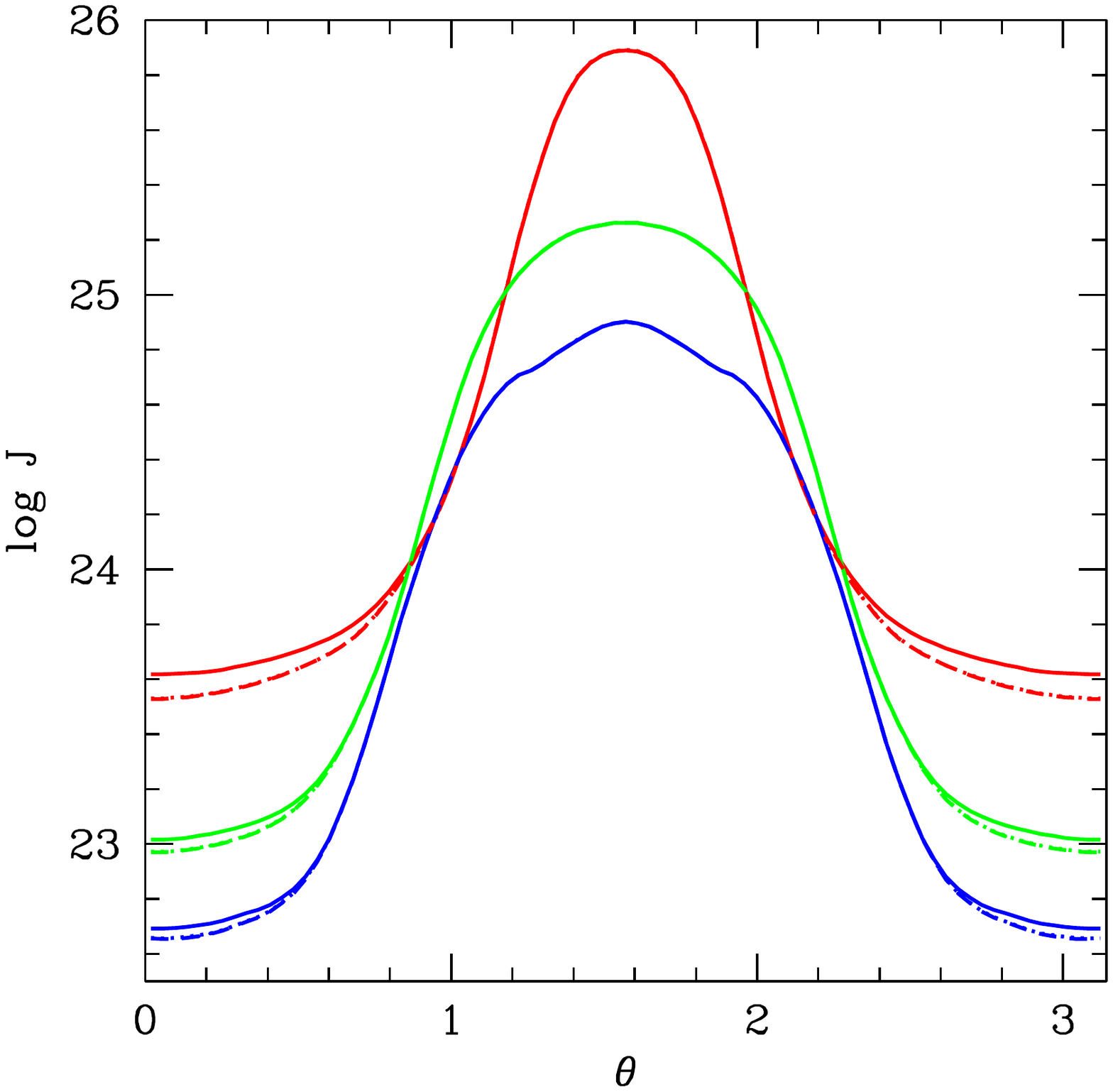}\hspace{-1.5cm}
\includegraphics[width=1.13\columnwidth]{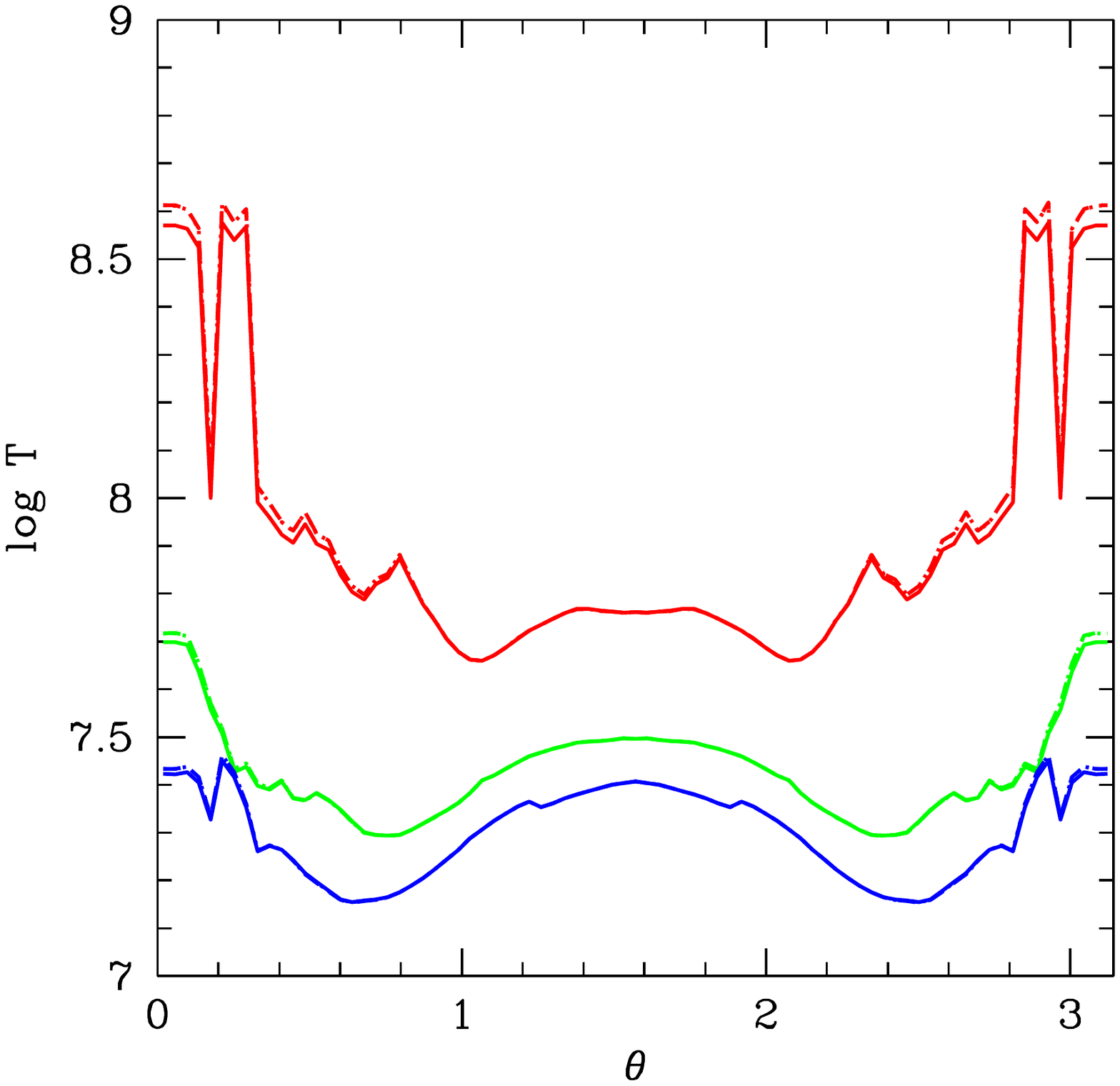} \\
\vspace{-2.5cm}
\caption{Top Left: Radiation intensity $J$ in the fiducial model
  r010\_3d as a function of Boyer-Lindquist angle $\theta$ at three
  Boyer-Lindquist radii: $r= 5M$ (red), $15M$ (green), $25M$
  (blue). Solid lines correspond to the \texttt{HEROIC} solution after
  Stage III (see Middle Left panel in
  Fig.~\ref{fig:jradint_temp}). Dotted lines correspond to the
  original solution from \texttt{KORAL} (Top Left,
  Fig.~\ref{fig:jradint_temp}).  Dashed lines correspond to the
  \texttt{HEROIC} solution after Stage I (Top Right,
  Fig.~\ref{fig:jradint_temp}). Top Right: Corresponding temperature
  profiles vs $\theta$ at the same three radii. Bottom Left: Profiles
  of $J$ in the \texttt{HEROIC} solution after Stage III (solid lines)
  and Stage II (dashed lines). Bottom Right: Corresponding temperature
  profiles.}
\label{fig:fid_jrad_temp}
\end{figure*}

The Middle Right panel in Figure~\ref{fig:jradint_temp} shows the
result we obtain after Stage III for the radiation intensity $J$ of
model r010\_2d, the 2D version of the fiducial model.  This
\texttt{KORAL} simulation ran significantly faster than the 3D model
(by a factor of tens), yet the results for the radiation field agree
surprisingly well with those of the 3D model shown in the Middle Left
panel (see also the comparison of spectra below). This suggests that
it is generally safe to use 2D models to compute radiation quantities
\citep{sadowski16a}. In the rest of the paper, we freely mix results
from 3D and 2D models. The one exception is the MAD models, which
cannot be run in 2D.

Figure \ref{fig:fid_jrad_temp} is another illustration of the same
results.  The top two panels show profiles of the mean radiation
intensity and the gas temperature as a function of polar angle
$\theta$. The different line types correspond to the original
\texttt{KORAL} solution (dotted lines), the \texttt{HEROIC} result
after Stage I, keeping the temperature fixed at the \texttt{KORAL}
values (dashed lines), and the \texttt{HEROIC} result after Stage III,
where the temperature is adjusted consistently (solid lines).  The
results generally confirm the previous discussion in connection with
Figure~\ref{fig:jradint_temp}. 
%That is, the \texttt{KORAL} radiation intensity is too low and the
%gas temperature is too high in the funnel, while the \texttt{HEROIC}
%solution for the radiation intensity without adjusting the
%temperature is too high.
The jagged structure of the temperature solutions is because the
viscous heating rate $q^+$ is patchy (Fig.~\ref{fig:extrapolate}). The
effect is most severe at small radii near the BH horizon and is mostly
restricted to moderately optically thick regions. It does not seem to
have a strong effect on observables.

The bottom two panels in Fig.~\ref{fig:fid_jrad_temp} compare two
\texttt{HEROIC} solutions. The solid lines show the standard solution
we have already described, in which the temperature is obtained
self-consistently and we apply both short characteristics and long
characteristics (Stage III). The dashed lines show the solution
obtained via the short characteristics method alone (Stage II), i.e.,
without doing a final round of long characteristics. The latter shows
only a small deviation from the more exact Stage III
calculation. Therefore, in principle, it may be sufficient to stop
after doing short characteristics, at least if we are interested only
in the radiation field at small radii. The differences are more
noticeable at larger radii because of the presence of ray defects in
the short characteristic solution (see \citealt{zhu15}).

We also computed a \texttt{HEROIC} solution of the fiducial model in
which we included thermal synchrotron radiation. The radiation field
and spectrum are virtually identical to those of the solution without
this emission. Thus, for the fiducial model at least, synchrotron
emission is negligible. This is not surprising, since the gas
temperature is below $10^9$\,K everywhere.

\subsection{Spectra and images}\label{fid_spectra}

\begin{figure*}
\includegraphics[width=1.13\columnwidth]{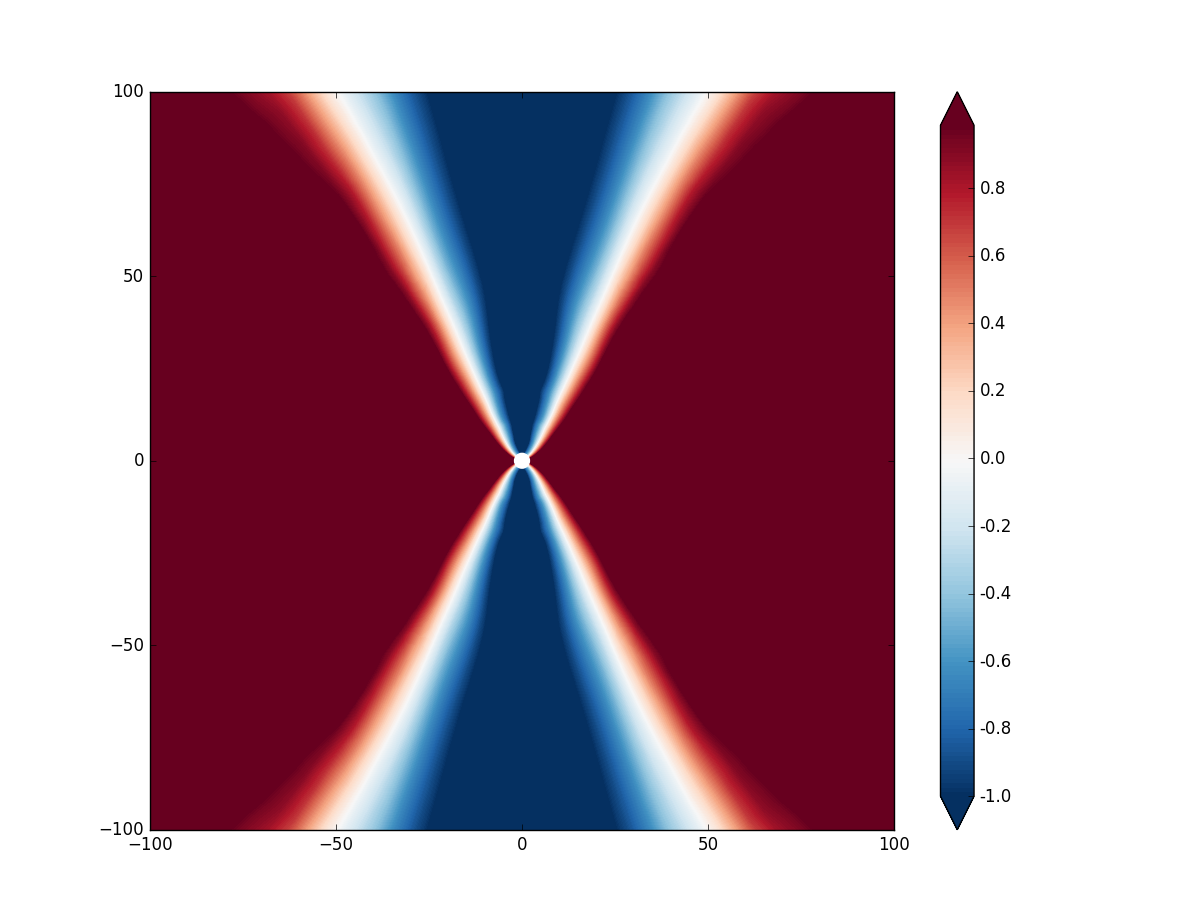}\hspace{-1.5cm}
\includegraphics[width=1.13\columnwidth]{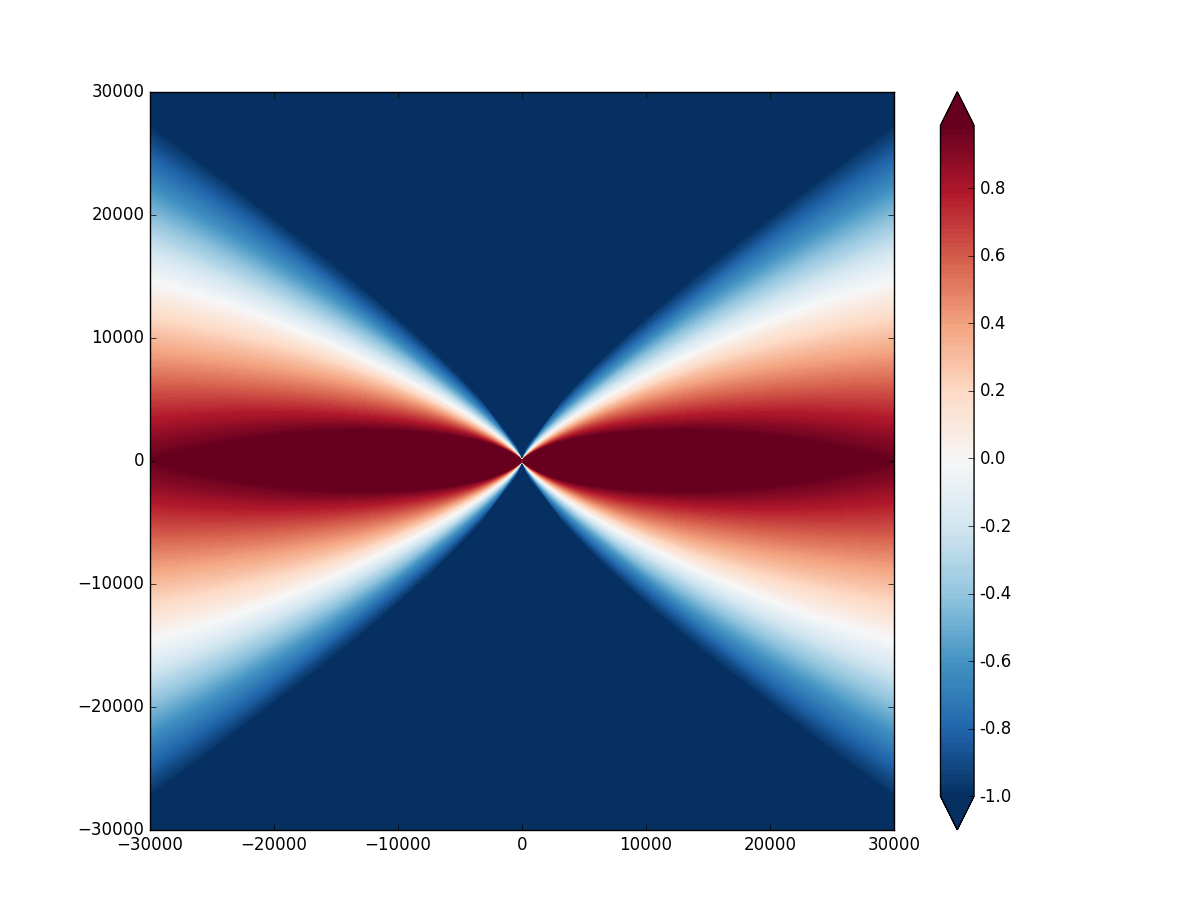} \\
\caption{Shows logarithm of the electron scattering optical depth
  $\log\tau$ as a function of position in the poloidal plane in the
  fiducial model r010\_3d.  Red regions are optically thick, blue
  regions are optically thin, and the white zone corresponds to
  $\tau\sim1$. The left panel shows a region close to the BH and the
  right panel corresponds to a larger region. The $\tau$ shown here is
  measured from the nearest pole.}
\label{fig:fid_tau}
\end{figure*}

Model r010\_3d has a large accretion rate of $10\dot{M}_{\rm Edd}$, so
the accretion flow is expected to be geometrically thick. This is
illustrated by the plots of the optical depth $\tau$ shown in
Fig.~\ref{fig:fid_tau}.  As we see, the optically thin funnel near the
BH has an opening angle less than $30^o$. Note that this angle is much
less than the funnel opening angle one might estimate from
Figure~\ref{fig:jradint_temp}. In fact, even Figure \ref{fig:fid_tau}
is a little misleading because $\tau$ here is measured from the pole,
at constant radius. If we instead computed the effective $\tau$ in the
radial direction to a distant observer, the funnel would appear even
narrower (as discussed later). The basic result, however, is the same,
namely, only observers within a fairly small angle of the pole are
able to see the intense radiation produced at the bottom of the
funnel. Observers at larger radii will still receive some radiation
from the walls of the funnel, plus of course emission from the disk
farther out, but the hottest region at the bottom of the funnel will
be invisible to them. The disk geometry thus has an obvious effect on
the observed spectrum as a function of inclination angle.

\begin{figure*}
\includegraphics[width=1.13\columnwidth]{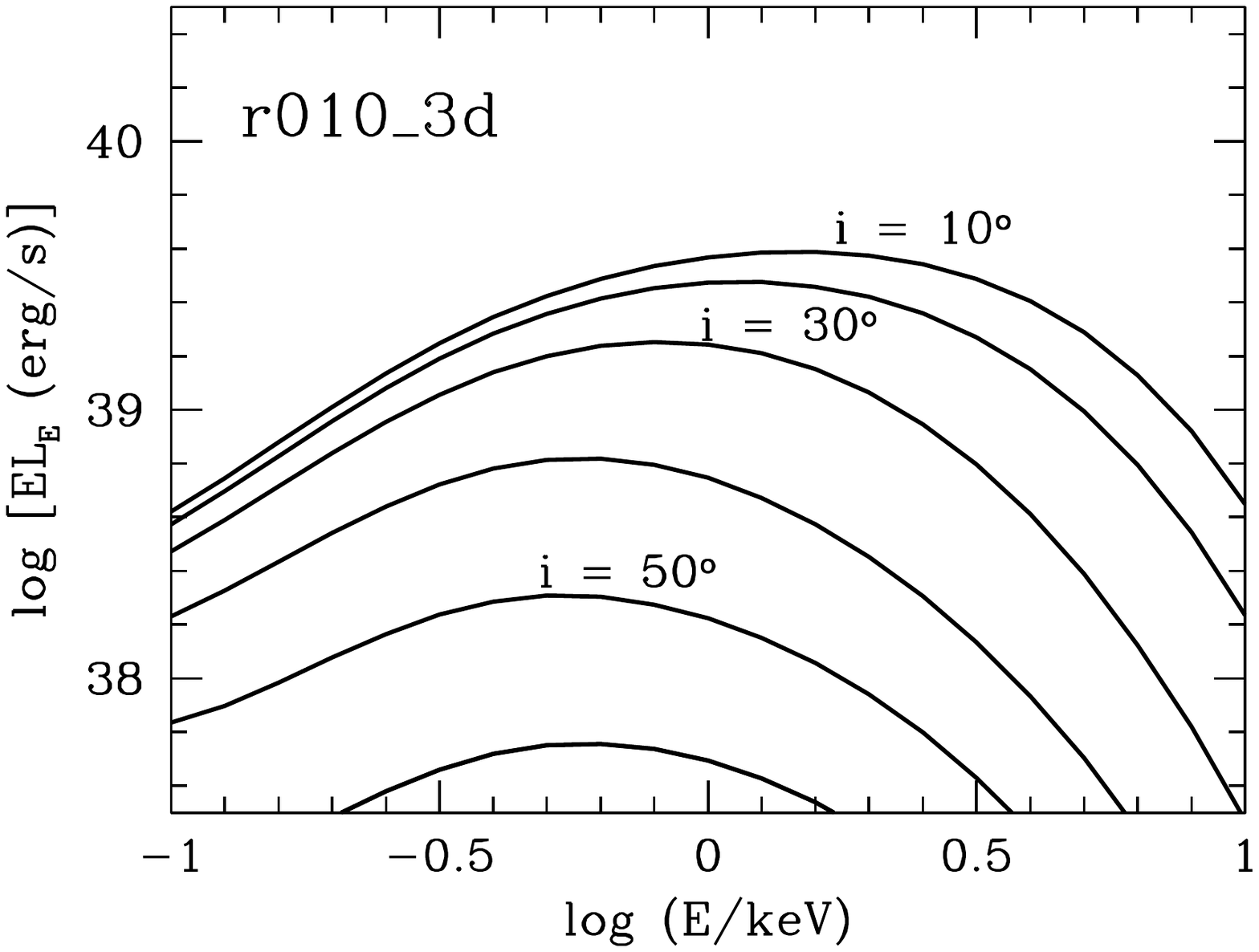}\hspace{-1.5cm}
\includegraphics[width=1.13\columnwidth]{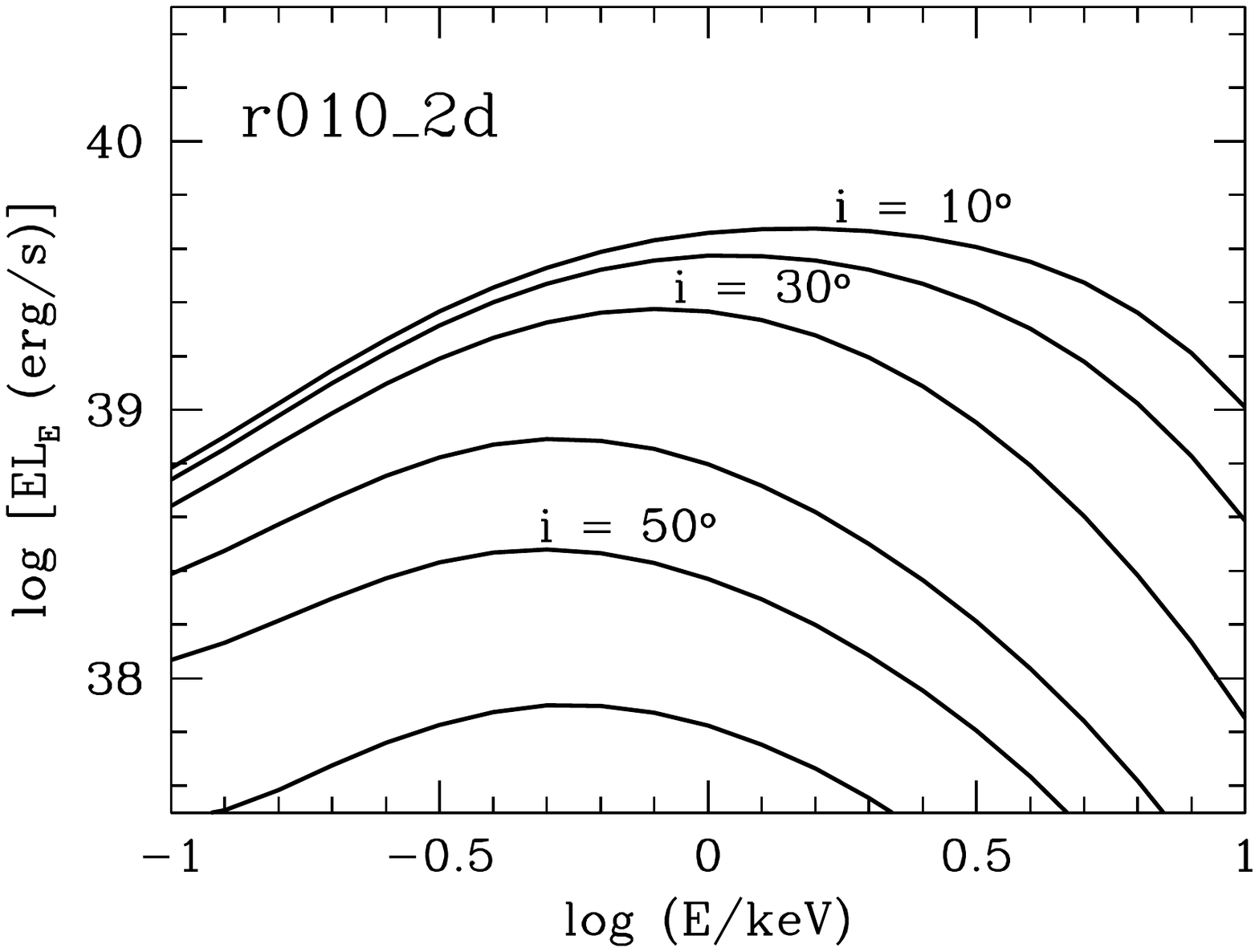} \\
\vspace{-4.5cm}
\caption{Left: Spectra of the fiducial model r010\_3d as seen by
  observers at different inclination angles $i$. Note the rapid
  softening of the spectrum and the fall-off of the luminosity for
  inclinations of $30$\,deg and above. Right: Shows spectra for the 2D
  model r010\_2d. The results are similar.}
\label{fig:fid_spectra}
\end{figure*}

Figure \ref{fig:fid_spectra} shows spectra computed by ray-tracing
(Stage IV) for observers at different inclination angles. For
inclination angles of $10^\circ$ and $20^\circ$, the observer sees a
fairly hot spectrum that peaks at several keV and has an isotropic
equivalent luminosity close to $10^{40}~{\rm erg\,s^{-1}}$. This is
fairly similar to spectra observed in ULXs.

Already at $i=30^\circ$, the most intense radiation from the bottom of
the funnel is no longer visible to the observer. The luminosity
decreases, and the spectrum softens dramatically. This effect becomes
more pronounced at higher inclinations. By $i=60^\circ$, the observed
luminosity is less than $10^{38}~{\rm erg\,s^{-1}}$. Interestingly,
this spectrum shows considerable resemblance to that of a ``classical"
supersoft source \citep{vandenheuvel92} or of an ultraluminous
supersoft source \citep{urquhart16}.

Figure \ref{fig:fid_spectra} also shows spectra computed from model
r010\_2d, which is the 2D version of the fiducial model. The computed
spectra are quite similar to those obtained from the 3D model, with a
small mismatch in the overall luminosity. This comparison is
encouraging, since 2D models are much cheaper to run than equivalent 3D
models.

\begin{figure*}
\includegraphics[width=1.13\columnwidth]{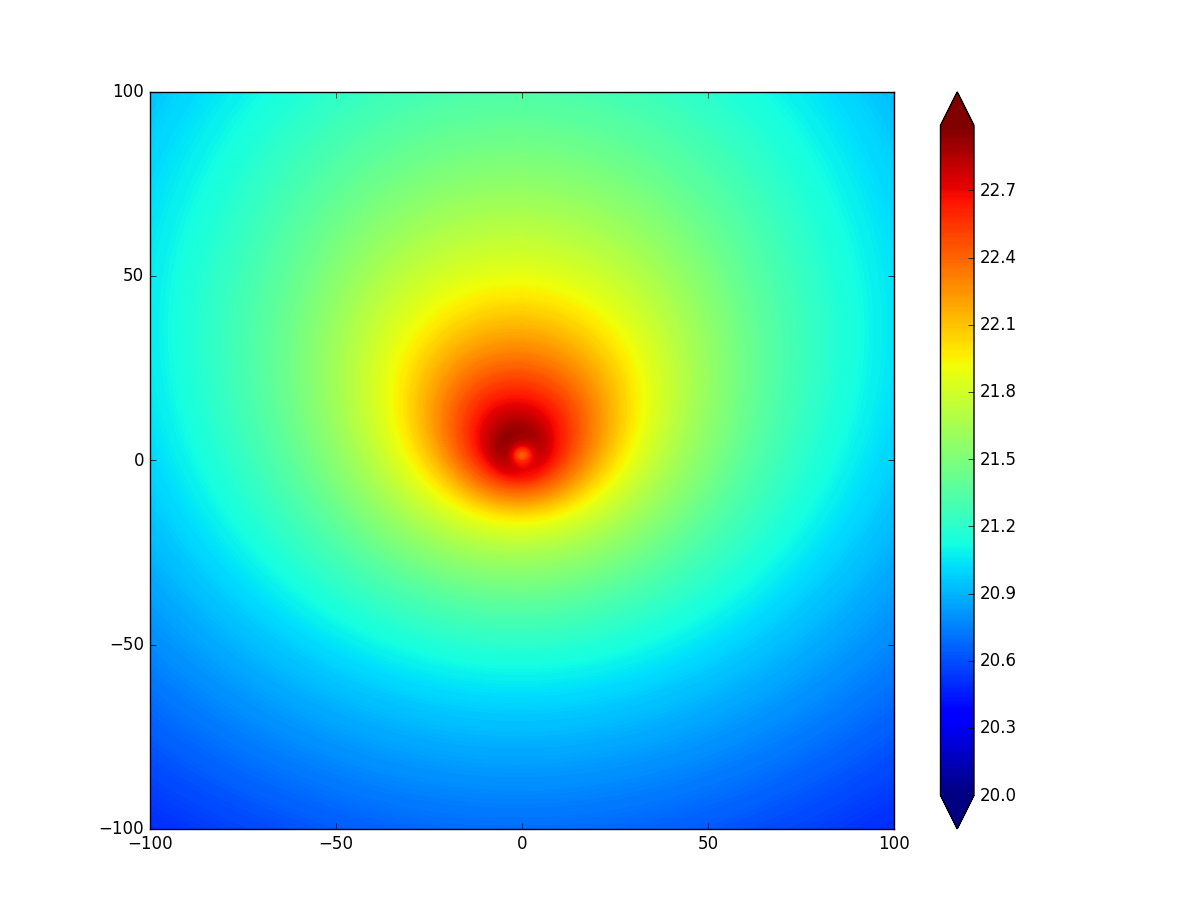}\hspace{-1.5cm}
\includegraphics[width=1.13\columnwidth]{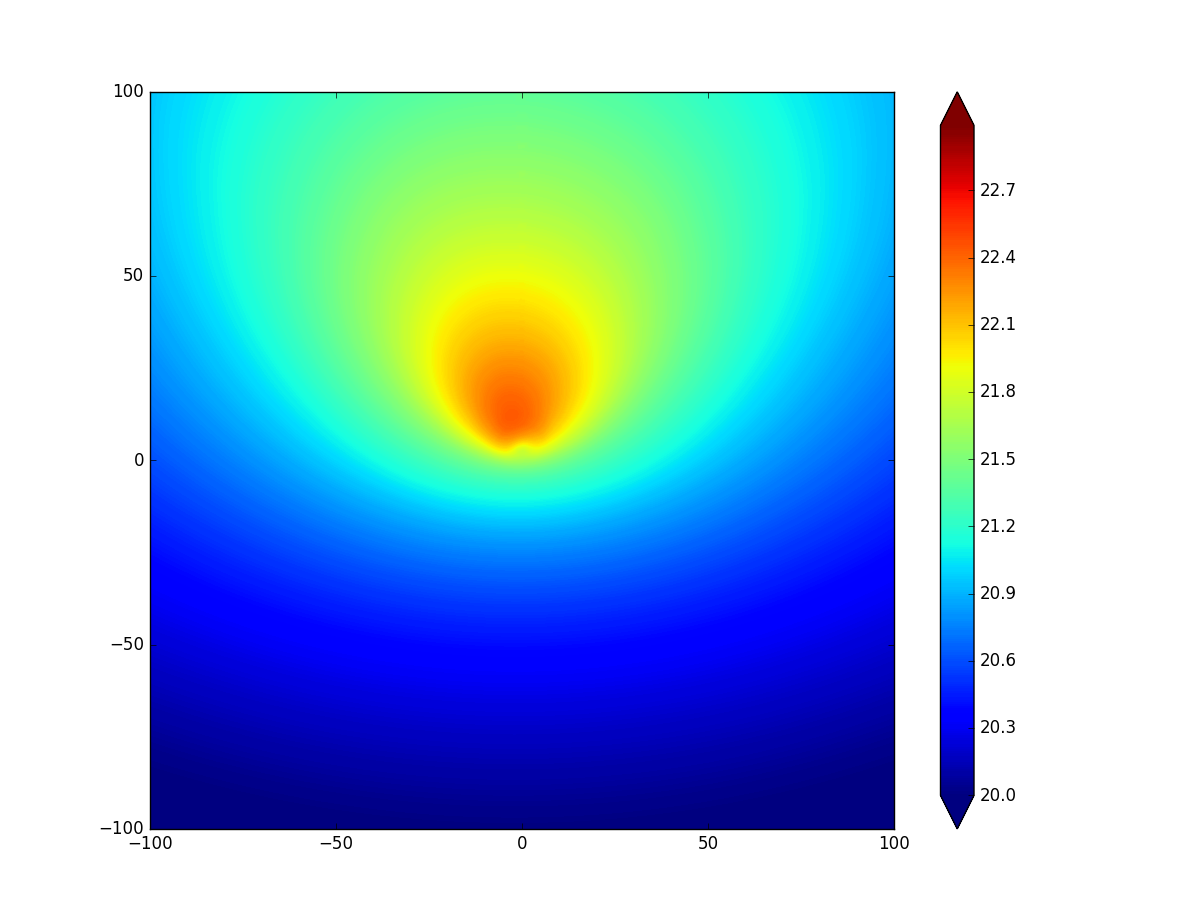} \\
\includegraphics[width=1.13\columnwidth]{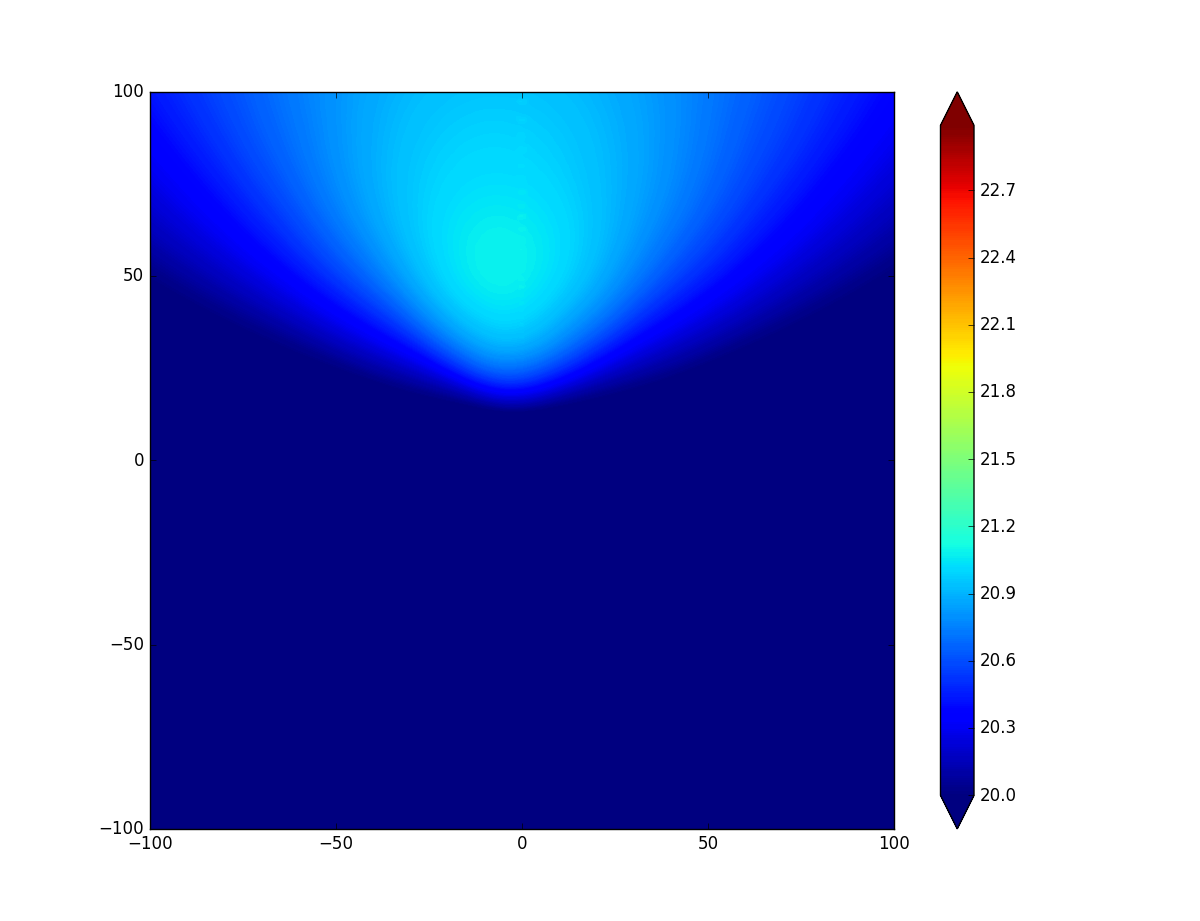}\hspace{-1.5cm}
\includegraphics[width=1.13\columnwidth]{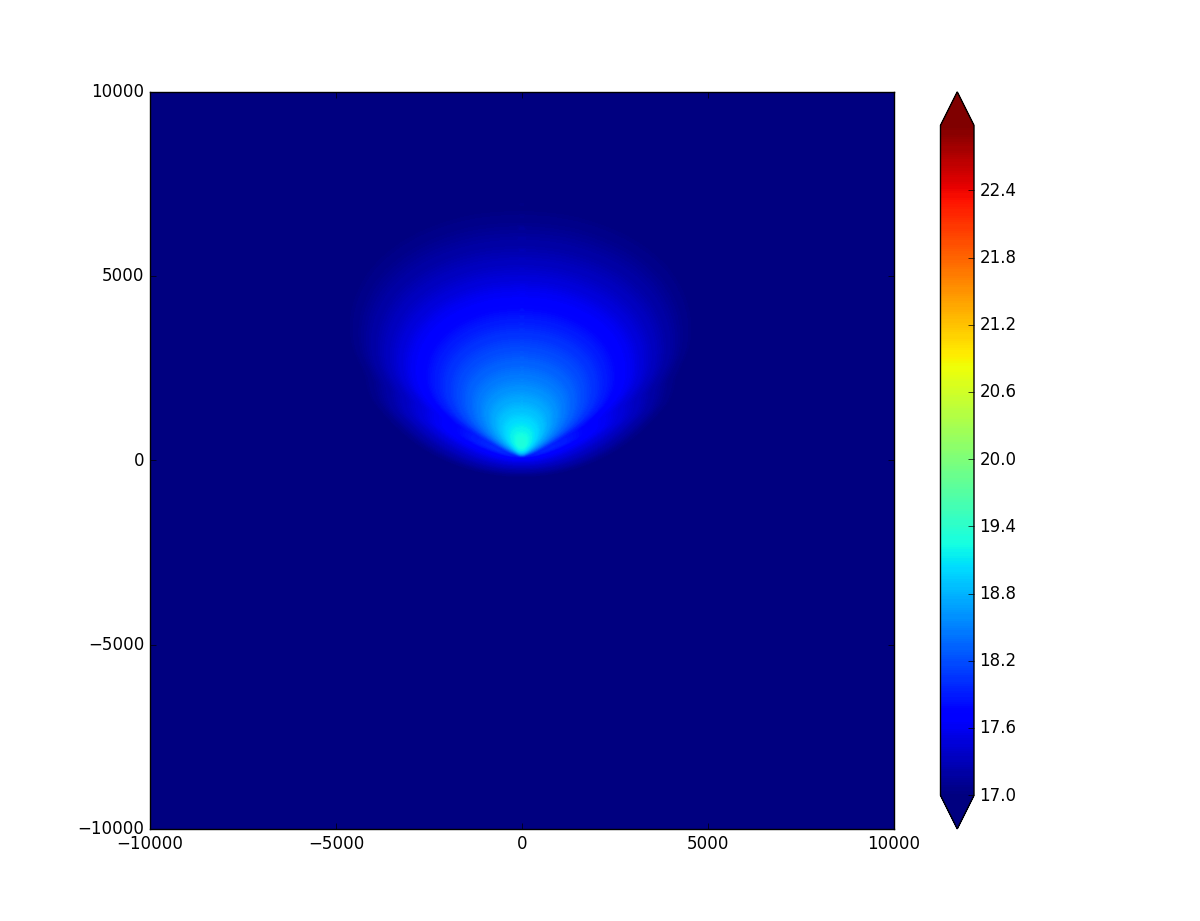} \\
\caption{Top Left: Image of the fiducial model r010\_3d as seen by an
  observer at inclination angle $i=10$\,deg from the pole. Color
  indicates logarithm of the brightness. Note the intense radiation
  emerging from near the BH, which dominates the observed
  radiation. Top Right: $i=20$\,deg. The bright region near the BH is
  still visible. Bottom Left: $i=30$\,deg. The bright region is no
  longer visible, and the observed radiation comes primarily from the
  wall of the funnel. Bottom Right: $i=40$\,deg. This is again
  dominated by radiation from the wall. Note that both the linear
  scale and the color scale have been expanded significantly in this
  panel.}
\label{fig:fid_images}
\end{figure*}

Fig~\ref{fig:fid_images} shows images computed with \texttt{HEROIC}
(Stage IV) for four inclination angles. They illustrate the
geometrical arguments that were used above to explain the dramatic
effect of the inclination angle on observed spectra.  As can be seen,
only observers at inclination angles $\leq 20^\circ$ receive radiation
from the hot bright region at the bottom of the funnel.  Already at
$30^\circ$, this region is hidden and the observed radiation is
dominated by emission from the funnel wall at tens of $GM/c^2$. By
$40^\circ$, the observer only sees regions of the funnel wall at large
radii. The observed luminosity drops rapidly and so does the spectral
hardness.

\section{Dependence on Parameters}\label{sec:parameters}

\subsection{Mass accretion rate}\label{sec:mdot}

\begin{table}
\begin{center}
\caption{Luminosities and efficiencies of models with $a_*=0$}
\label{tab:luma0}
\begin{tabular}{lccccc}
\hline
\hline
Model & $\dot{M}$$^1$ & $L_{\rm rad}$$^2$ & $L_{\rm total}$$^2$ & 
$\eta_{\rm rad}$ & $\eta_{\rm total}$\\
\hline
\hline
\texttt{r010\_3d/SANE} & 10 & 2.0 & 6.6 & 0.009 & 0.030\\
\texttt{r010\_2d/SANE} & 10 & 2.3 & 9.4 & 0.010 & 0.043\\
\texttt{r012\_3d/SANE} & 1.2 & 0.7 & 1.2 & 0.027 & 0.044\\
\texttt{r030\_2d/SANE} & 7.0 & 2.4 & 5.7 & 0.016 & 0.037\\
\texttt{r031\_2d/SANE} & 17 & 2.4 & 16 & 0.006 & 0.043\\
\hline
\texttt{r013\_3d/MAD} & 23 & 1.6 & 18 & 0.003 & 0.035\\
\texttt{r023\_3d/MAD} & 1.3 & 0.76 & 3.2 & 0.027 & 0.111\\
\hline
\hline
\end{tabular}
\end{center}
$^1$ In units of $\dot{M}_{\rm Edd}$ with $\eta_{\rm NT} = 0.05719$ \\
$^2$ In units of $10^{39}\,{\rm erg\,s^{-1}}$ 
\end{table}

\begin{figure}
\begin{center}
\includegraphics[width=0.5\textwidth]{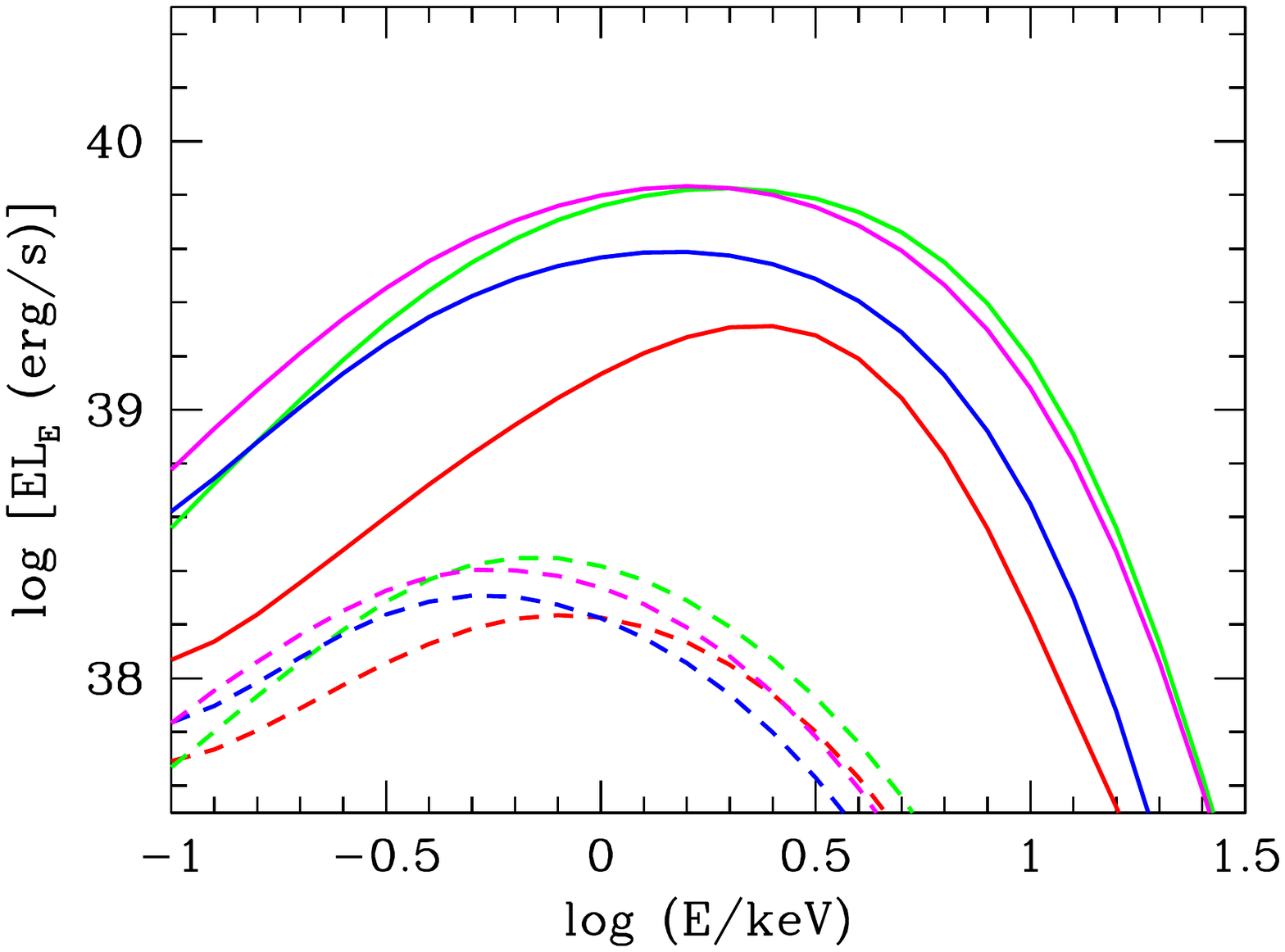}
\vspace{-4.5cm}
\caption{Spectra of models with $M=10M_\odot$, $a_*=0$, SANE magnetic
  field, and varying $\dot{M}$. Colors are as follows: $\dot{M} =
  1.2\dot{M}_{\rm Edd}$ (red, model r012\_3d)), $\dot{M} = 7\dot{M}_{\rm
    Edd}$ (green, model r030\_2d), $\dot{M} = 10\dot{M}_{\rm Edd}$
  (blue, model r010\_3d), $\dot{M} = 17\dot{M}_{\rm Edd}$ (magenta,
  model r031\_2d). Solid lines correspond to an observer inclination
  angle $i=10^\circ$, and dashed lines to $i=50^\circ$.  }
\label{fig:a0_mdot}
\end{center}
\end{figure}

Figure \ref{fig:a0_mdot} shows the effect of changes in the mass
accretion rate. The four models have $M=10M_\odot$, $a_*=0$, SANE
magnetic field, and $\dot{M} = 1, ~7, ~10, ~17\dot{M}_{\rm Edd}$,
respectively. For an observer inclination angle of $10^\circ$, model
r012\_3d, with the lowest $\dot{M}=1\dot{M}_{\rm Edd}$, has a spectrum
not unlike that of a thin accretion disk model.  The three other
models behave differently, with their spectra showing a much broader
peak. In fact, these three models are quite similar to one another,
both in luminosity and spectrum, which suggests that at higher
$\dot{M}$, the observed spectrum is insensitive to the accretion
rate. Curiously, model r010\_3d is less luminous than models r030\_2d
and r031\_2d, even though its mass accretion rate lies in between the
other two models. This is in part the result of a general trend we
see, namely, that 3D simulations with \texttt{KORAL} tend to be a
little less luminous than 2D models with the same parameters.

At an inclination angle of $50^\circ$, all four models in Figure
\ref{fig:a0_mdot} have substantially lower luminosity and have much
softer spectra. This is because the models are sufficiently
geometrically thick --- even in the case of the $1\dot{M}_{\rm Edd}$
model --- that the inner region of the disk is screened from the view
of the observer.  Thus, the observer sees only cooler and less
luminous radiation from larger radii.

While the above discussion is related to specific observer
inclinations, Table~\ref{tab:luma0} shows the total radiative
luminosities $L_{\rm rad}$ of the $a_*=0$ models integrated over all
angles.  These show the same pattern as a function of $\dot{M}$. A
particularly striking result is that $L_{\rm rad}$ apparently
saturates at roughly $2L_{\rm Edd}$, even for quite super-Eddington
accretion rates. Also shown in Table~\ref{tab:luma0} is the radiative
efficiency $\eta_{\rm rad}$, defined as
\begin{equation}
\eta_{\rm rad} \equiv \frac{L_{\rm rad}}{\dot{M}c^2}.
\label{eq:etarad}
\end{equation}
The high-$\dot{M}$ models are clearly radiatively inefficient.  This
is expected for the super-Eddington ``slim disk'' \citep{abramowicz88}
regime of accretion, where advection dominates.

In contrast to the radiative luminosity, the mechanical energy output
of slim disks via jets and winds is not Eddington-limited. This is
reflected in Table~\ref{tab:luma0} in the total luminosities $L_{\rm
  total}$ (radiation+jet+wind) and the corresponding efficiencies,
\begin{equation}
\eta_{\rm total} \equiv \frac{L_{\rm total}}{\dot{M}c^2}.
\label{eq:etatotal}
\end{equation}
Note in particular the high $\dot{M}$ model r031\_2d, which has a
radiative luminosity of only $2.4\times10^{39}~{\rm erg\,s^{-1}}$ and
a corresponding radiative efficiency of only 0.6\%, whereas its total
luminosity and total efficiency are $1.6\times10^{40}~{\rm
  erg\,s^{-1}}$ and 4.3\%.

\subsection{Black hole spin}\label{sec:spin}

\begin{table}
\begin{center}
\caption{Luminosities and efficiencies of models with $a_*=0.9$}
\label{tab:luma9}
\begin{tabular}{lccccc}
\hline
\hline
Model & $\dot{M}$$^1$ & $L_{\rm rad}$$^2$ & $L_{\rm total}$$^2$ 
& $\eta_{\rm rad}$ & $\eta_{\rm total}$$^3$\\
\hline
\hline
\texttt{r011\_3d/SANE} & 12 & 2.2 & 9.1 & 0.023 & 0.094 \\
\texttt{r032\_2d/SANE} & 6.2 & 2.0 & 4.3 & 0.039 & 0.087\\
\texttt{r033\_2d/SANE} & 10 & 2.5 & 6.6 & 0.031 & 0.082\\
\texttt{r034\_2d/SANE} & 26 & 1.7 & 24 & 0.008 & 0.116\\
\hline
\texttt{r014\_3d/MAD} & 36 & 57 & 200 & 0.20 & 0.69\\
\texttt{r015\_3d/MAD} & 6.8 & 7.2 & 39 & 0.13 & 0.71\\
\hline
\hline
\end{tabular}
\end{center}
$^1$ In units of $\dot{M}_{\rm Edd}$ with $\eta_{\rm NT} = 0.1558$ \\
$^2$ In units of $10^{39}\,{\rm erg\,s^{-1}}$ 
\end{table}

\begin{figure}
\begin{center}
\includegraphics[width=0.5\textwidth]{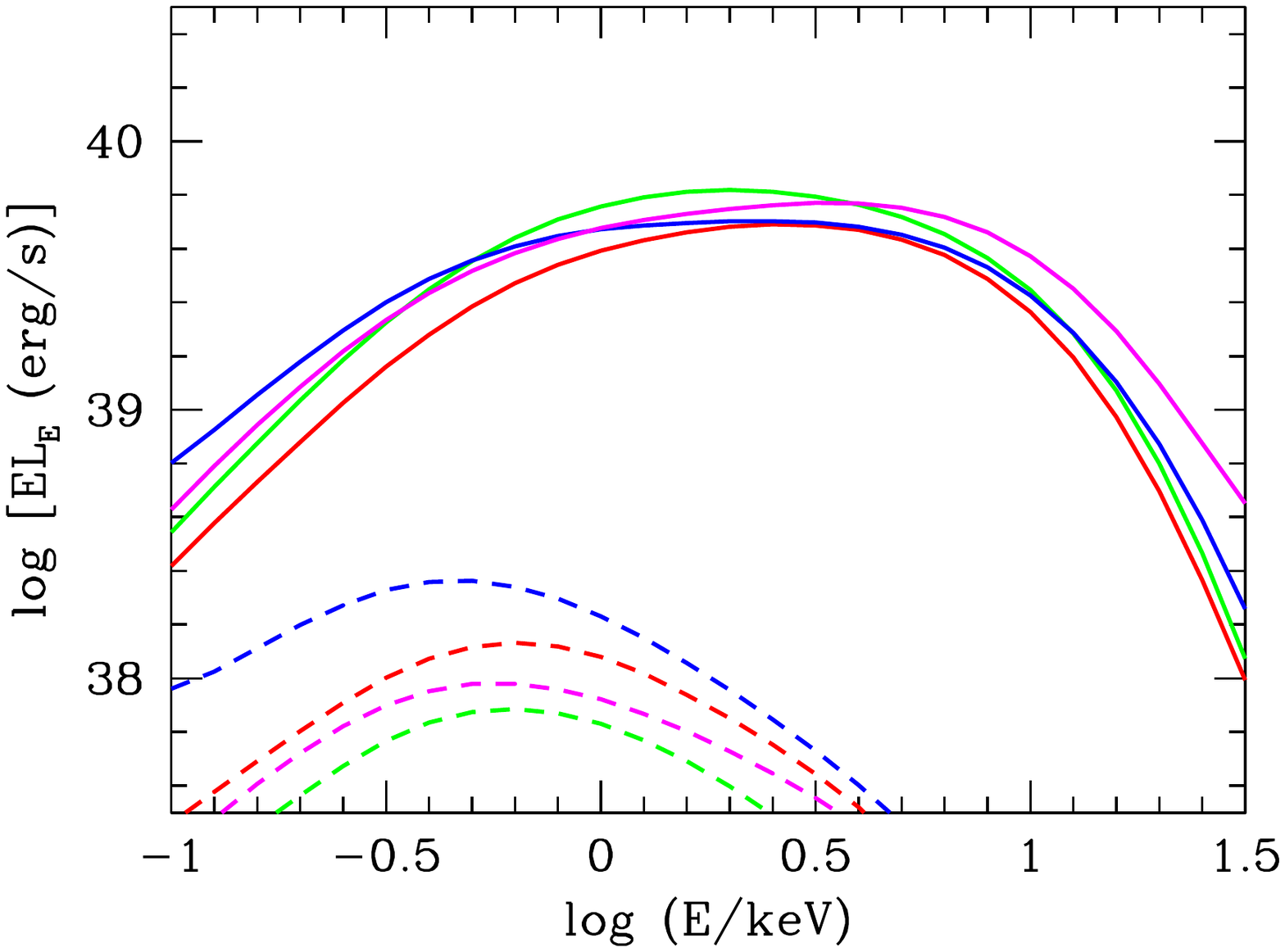}
\vspace{-4.5cm}
\caption{Spectra of models with $M=10M_\odot$, $a_*=0.9$, SANE
  magnetic field, and varying $\dot{M}$. Colors are as follows:
  $\dot{M} = 6.2\dot{M}_{\rm Edd}$ (red, model r032\_2d), $\dot{M} =
  10\dot{M}_{\rm Edd}$ (green, r033\_2d), $\dot{M} = 12\dot{M}_{\rm
    Edd}$ (blue, r011\_2d), $\dot{M} = 26\dot{M}_{\rm Edd}$
  (magenta, r034\_2d). Solid lines correspond to an observer
  inclination angle $i=10^o$, and dashed lines to $i=50^o$. }
\label{fig:a9_mdot}
\end{center}
\end{figure}

Figure \ref{fig:a9_mdot} is similar to Figure \ref{fig:a0_mdot},
except that the models considered here have spin $a_*=0.9$. These
spectra have the same general shape as the $a_*=0$ models. However,
the spectra are noticeably harder. This suggests that it might be
possible to obtain a rough estimate of the BH spin from spectral
hardness. However, the method works only for observers at favorable
inclination angles. For larger inclinations, the spectra are soft and
are not very different from those of the $a_*=0$ models. As in
Figure~\ref{fig:a0_mdot}, the spectra in Figure~\ref{fig:a9_mdot} are
again insensitive to $\dot{M}$.

Table~\ref{tab:luma9} lists the radiative and total luminosities and
corresponding efficiencies of the $a_*=0.9$ models. The pattern is
similar to that shown by the $a_*=0$ models. The radiative luminosity
saturates at a couple of Eddington, which implies that the radiative
efficiency falls substantially with increasing $\dot{M}$; for
instance, $\eta_{\rm rad}$ is only 0.8\% for modelr034\_2d. In
contrast, the total efficiency is around 10\%, independent of
$\dot{M}$.

\subsection{MAD models}\label{sec:mad}

\begin{figure*}
\includegraphics[width=1.13\columnwidth]{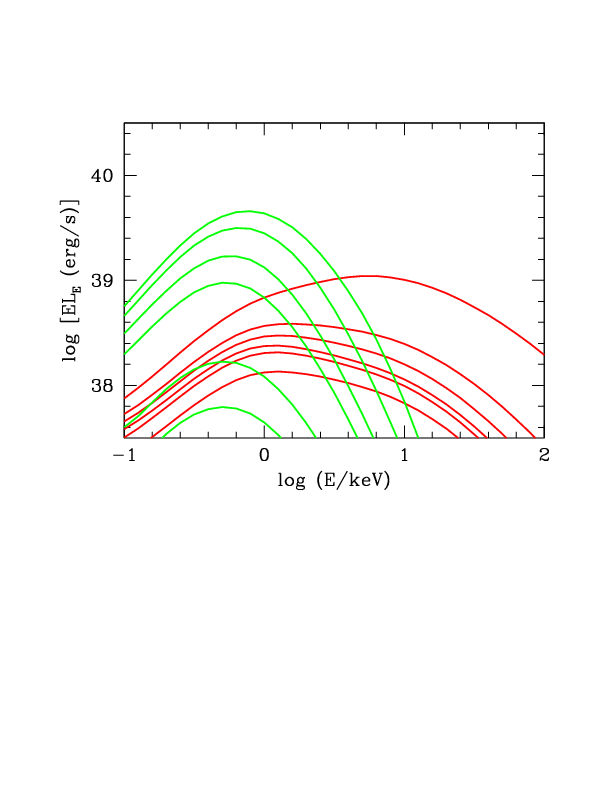}\hspace{-1.5cm}
\includegraphics[width=1.13\columnwidth]{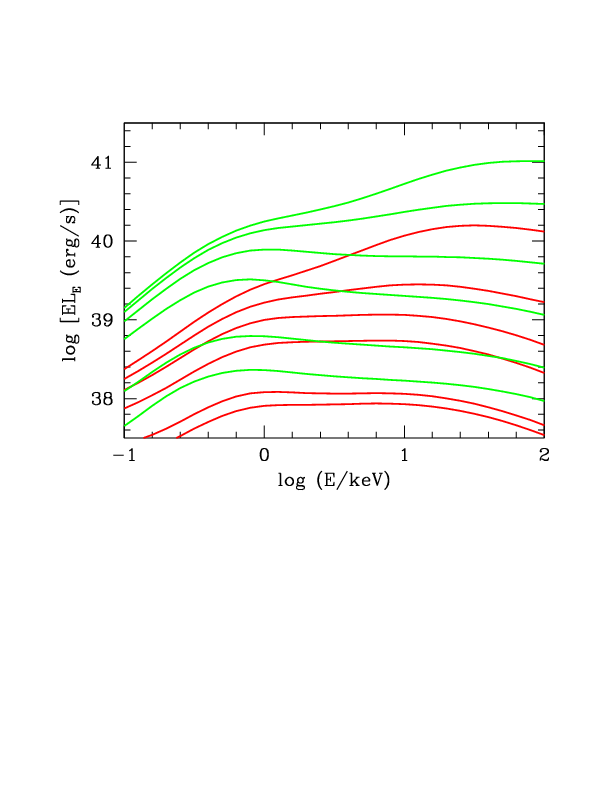} \\
\vspace{-4.5cm}
\caption{Left: Spectra of two MAD models with BH spin $a_*=0$. The
  different-colored lines correspond to $\dot{M} = 1.3\dot{M}_{\rm Edd}$
  (red, model r023\_3d) and $23\dot{M}_{\rm Edd}$ (green, model
  r013\_3d), respectively. For a given color, from above, the lines
  are for different observer inclinations: $i=10^\circ$, $20^\circ$,
  $30^\circ$, $40^\circ$, $50^\circ$, $60^\circ$. Right: MAD models
  with $a_* = 0.9$, and $\dot{M} = 6.8\dot{M}_{\rm Edd}$ (red, model
  r015\_3d) and $36\dot{M}_{\rm Edd}$ (green, model r014\_3d),
  respectively.}
\label{fig:mad}
\end{figure*}

\begin{figure*}
\includegraphics[width=0.45\columnwidth]{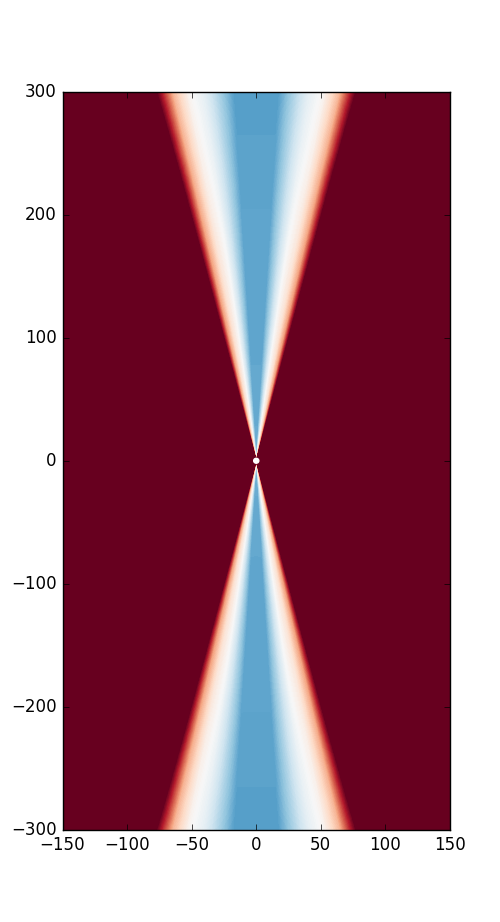}\hspace{0cm}
\includegraphics[width=0.45\columnwidth]{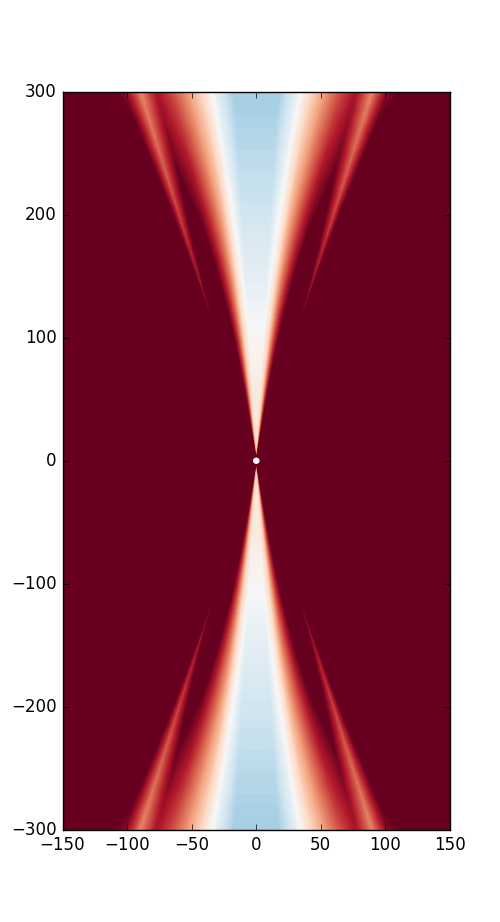}
\includegraphics[width=0.45\columnwidth]{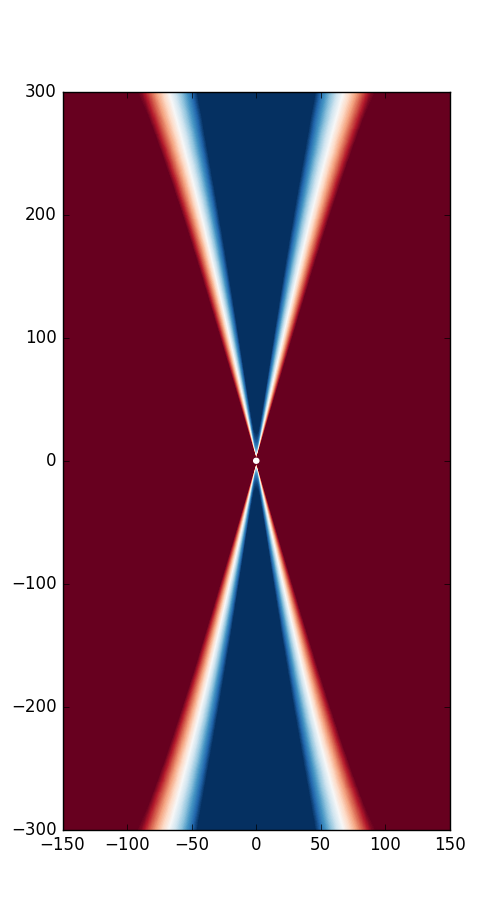}
\includegraphics[width=0.54\columnwidth]{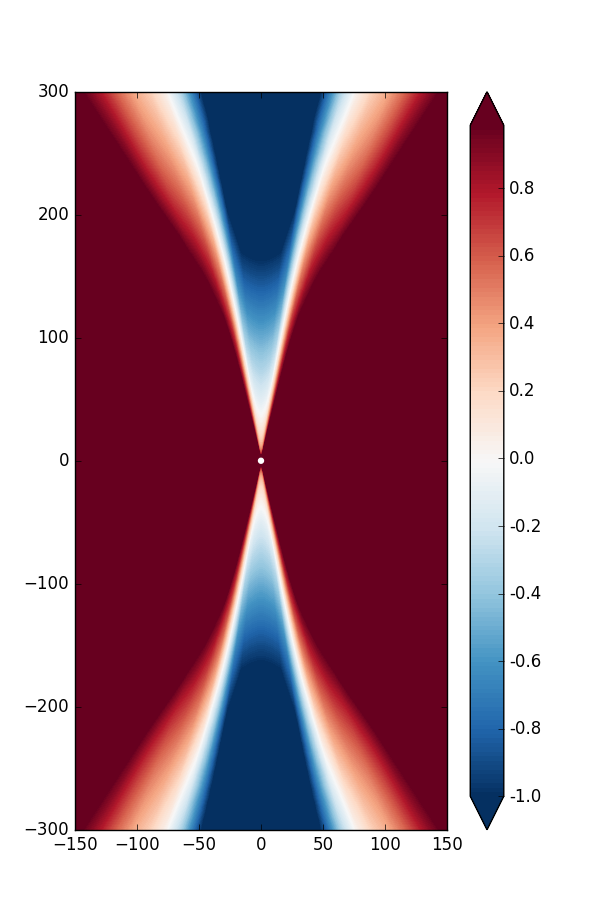} \\
\includegraphics[width=0.45\columnwidth]{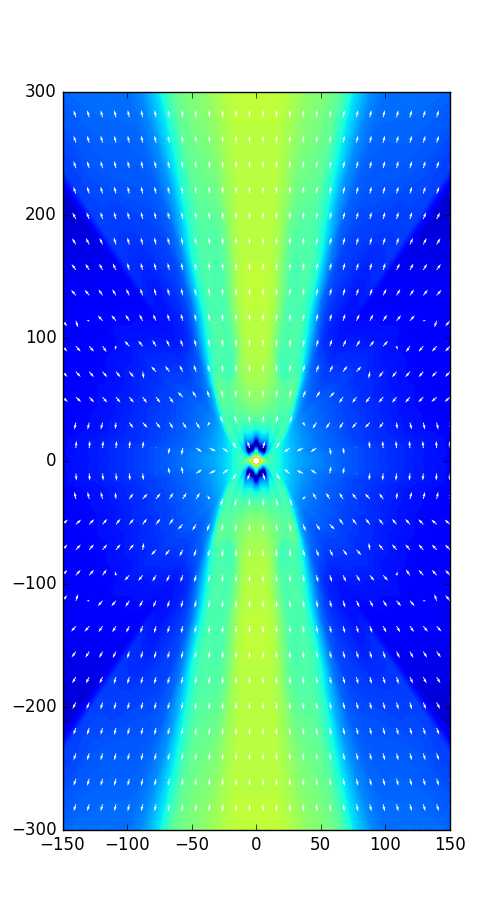}\hspace{0cm}
\includegraphics[width=0.45\columnwidth]{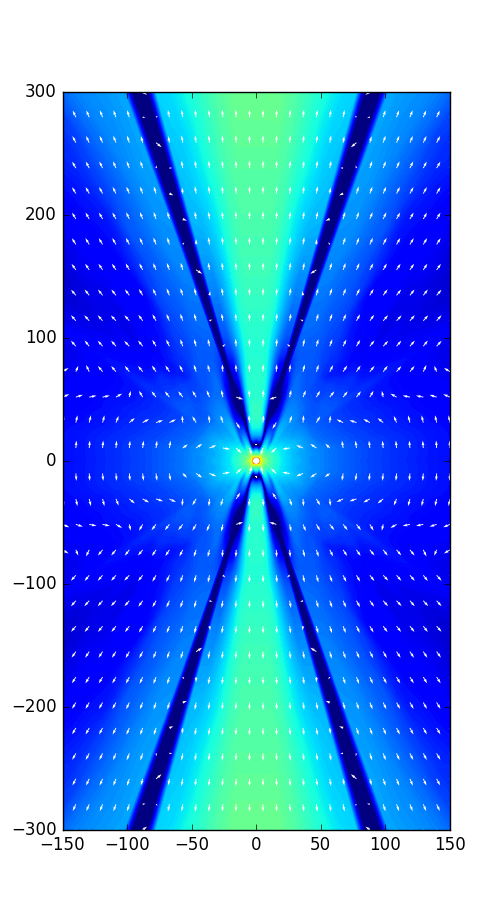}
\includegraphics[width=0.45\columnwidth]{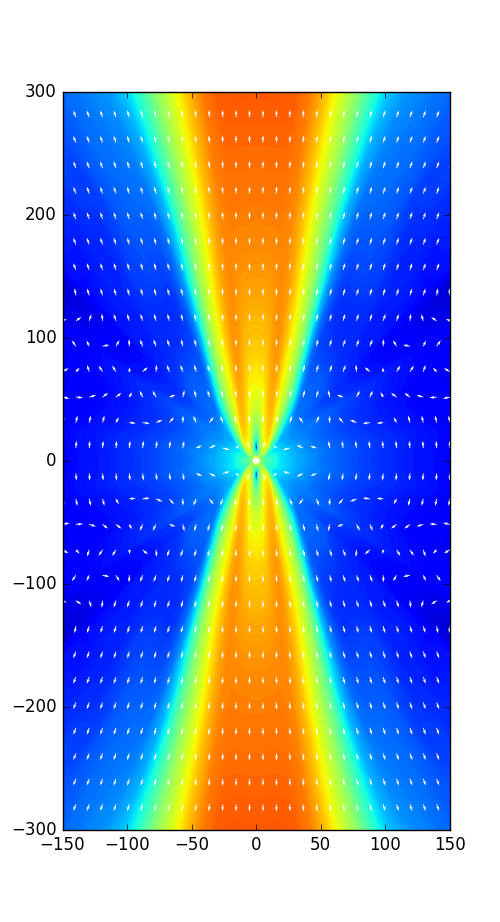}
\includegraphics[width=0.54\columnwidth]{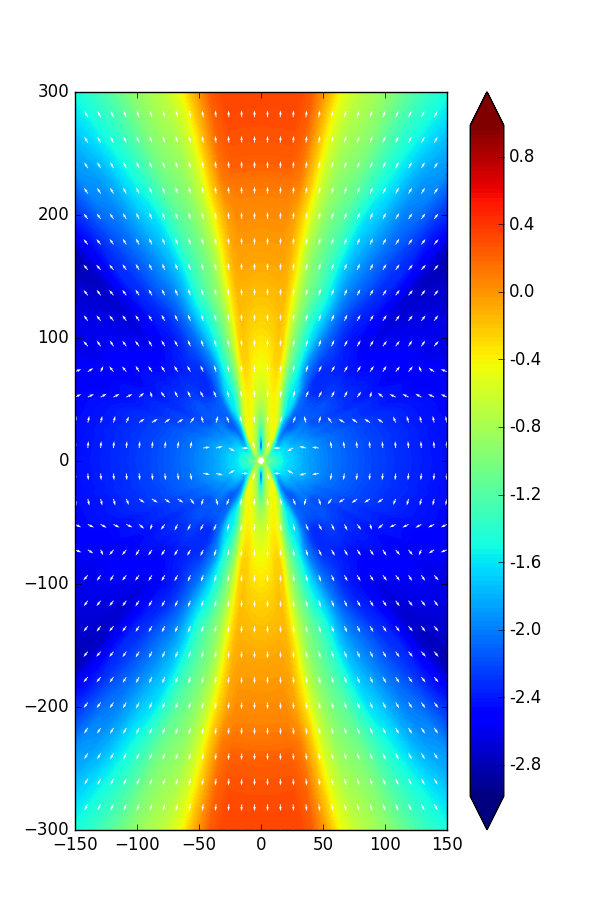} \\
\vspace{0cm}
\caption{Top Panels: Logarithm of the optical depth $\log\tau_{\rm
    radial}$ for four MAD models: Left: $a_*=0$, $\dot{M} =
  1.3\dot{M}_{\rm Edd}$ (model r023\_3d), Left Center: $a_*=0$, $\dot{M}
  = 23\dot{M}_{\rm Edd}$ (model r013\_3d), Right Center: $a_*=0.9$,
  $\dot{M} = 6.8\dot{M}_{\rm Edd}$ (model r015\_3d), Right: $a_*=0.9$,
  $\dot{M} = 36\dot{M}_{\rm Edd}$ (model r014\_3d). Bottom Panels:
  Shows the quantity $(\gamma-1)$ (color scale), where $\gamma$ is the
  bulk Lorentz factor of the gas. The arrows indicate the direction of
  motion of the gas in the poloidal plane.}
\label{fig:mad_tau}
\end{figure*}

The four MAD models listed in Table \ref{tab:models} are all run in
3D. In these models, the magnetic field near the BH and in the inner
region of the accretion disk is very strong, so much so that direct
accretion via an axisymmetric disk is not possible. Gas can accrete
only via non-axisymmetric streams and blobs, triggered by the
interchange (or other similar) instability
\citep{igumenshchev03,narayan03,tchekhovskoy11,mckinney12,mckinney15}. Since
the presence of a non-axisymmetric flow is a key feature of the MAD
regime, MAD models have to be run in 3D and are quite expensive.  We
report results here for four MAD models.

Figure \ref{fig:mad} shows spectra of the four models. The two models
with spin 0 (Left panel) display an unusual behavior: the model with a
higher $\dot{M}=23\dot{M}_{\rm Edd}$ has a very much softer spectrum
than the one with a lower $\dot{M} = 1.3\dot{M}_{\rm Edd}$. The reason
for this unexpected behavior can be understood from
Figure~\ref{fig:mad_tau}. The left two panels in the top row show the
optical depth $\tau_{\rm radial}$ as measured along the radius from
infinity (this is different from the $\tau$ shown in
Fig.~\ref{fig:fid_tau}). Notice that the funnel in the $23\dot{M}_{\rm
  Edd}$ model is optically quite thick.  This means that even
observers who are perfectly aligned with the axis do not receive
radiation directly from the bottom of the funnel, but rather from a
photosphere at a large radius $\sim100GM/c^2$.  Correspondingly, the
received radiation tends to be very soft.  The $1.3\dot{M}_{\rm Edd}$
model has less opacity in the funnel, so an aligned observer can see
farther down into the funnel and observes a harder spectrum.

The above discussion is fairly specific to non-spinning (or
slowly-spinning) BHs. Because of the lack of (or at best weak)
frame-dragging, these systems do not have an extra power source in the
BH ergosphere, as needed for the \citet{bz77} mechanism of powering
jets. The primary power source is the accretion disk. Any radiation
that flows into the funnel carries some gas with it, thereby enhancing
the opacity in the funnel.

The two MAD models with spin $a_*=0.9$ are quite different.  These
models are substantially more luminous and also have very hard spectra
(Fig.~\ref{fig:mad}). The models are highly jet-dominated, as is
evident from Figure~\ref{fig:mad_tau}. The jets clearly receive their
power from the spinning BH via the \citet{bz77} mechanism (which can
be understood as a generalized version of the \citealt{penrose69}
process, see \citealt{lasota14}). The funnels in these models are
quite empty of gas (even in the case of the model with
$\dot{M}=36\dot{M}_{\rm Edd}$), presumably because the gas is rapidly
blown away by the powerful jet.  As a result, observers at low
inclination angles can see down to the base of the funnel.

The lower panels in Figure \ref{fig:mad_tau} show the bulk Lorentz
factor of the gas in the funnel for the four MAD models. The two
$a_*=0$ models have gas with only modest velocities, whereas the two
$a_*=0.9$ models show quite relativistic motions. The latter models
have powerful jets driven by the BH spin. Relativistic beaming thus
causes both the luminosity and the spectral hardness to be strongly
enhanced. In fact, model r014\_3d has an apparent luminosity
$>10^{41}\,{\rm erg\,s^{-1}}$ for an observer at inclination angle
$10^\circ$, which corresponds to $>100L_{\rm Edd}$ for the given BH
mass.

Tables~\ref{tab:luma0} and \ref{tab:luma9} list the luminosities and
efficiencies of the MAD models. The $a_*=0$ MAD models have similar
luminosities as their SANE counterparts. The $a_*=0.9$ MAD models, on
the other hand, are substantially more luminous than equivalent SANE
models.

\subsection{Dependence of Luminosity on Parameters}\label{sec:luminosity}

\begin{figure}
\includegraphics[width=1.05\columnwidth]{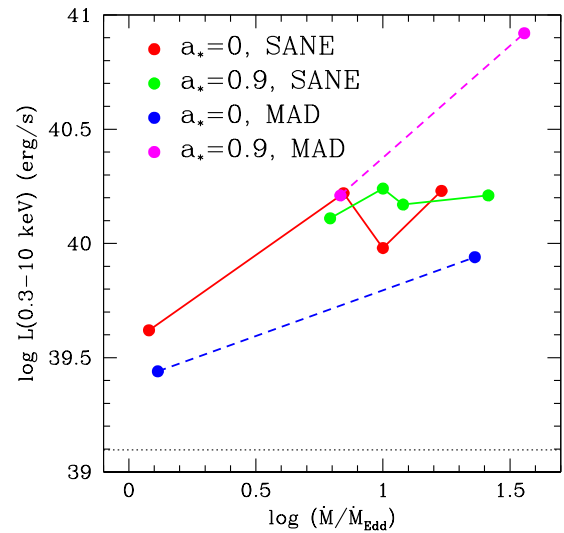}
\vspace{-3cm}
\caption{Apparent luminosity in the $0.3-10$\,keV band for an observer
  at inclination angle 10\,deg, as a function of the mass accretion
  rate $\dot{M}$, for sequences of SANE and MAD models with BH spin
  values of $a_*=0$ and 0.9. The horizontal dotted line corresponds to
  the Eddington luminosity.}
\label{fig:2kev_mdot}
\end{figure}

Figure \ref{fig:2kev_mdot} shows the apparent luminosity in the
$(0.3-10)$\,keV band as a function of $\dot{M}$ for the various model
sequences. The observer is assumed to be at an inclination angle of
$10^\circ$. SANE models appear to saturate in luminosity at about
$2\times10^{40} ~{\rm erg\,s^{-1}}$ (for a BH mass of $10M_\odot$).
MAD models show much more variation. Especially when the BH is
spinning, MAD models can be extremely luminous.
% (note their bolometric
%luminosities are a factor of several larger than their $0.3-10$\,keV
%luminosities because of their very hard spectra, see
%Fig.~\ref{fig:mad}, right panel).

\begin{figure*}
\includegraphics[width=1.05\columnwidth]{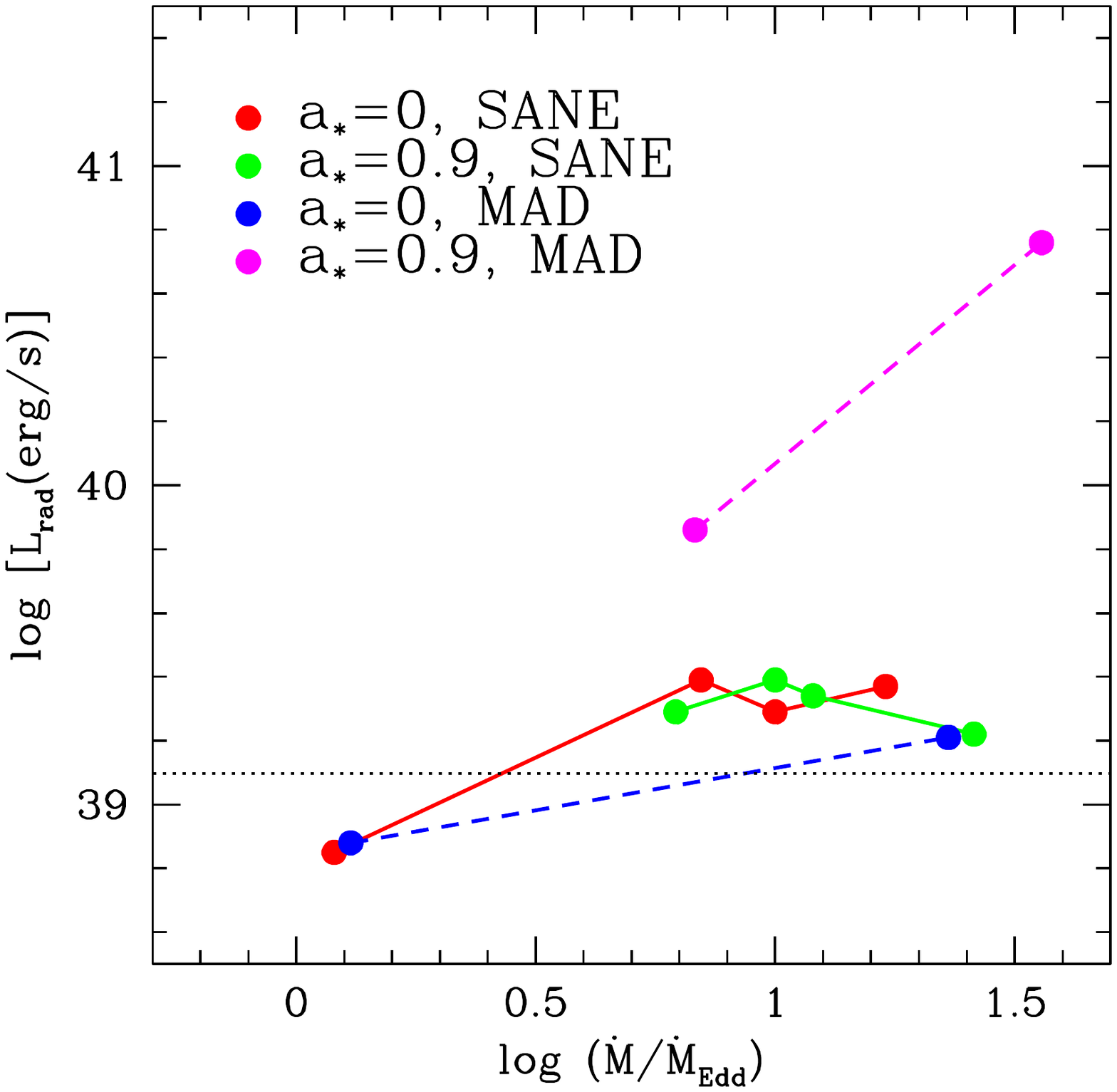}\hspace{-1cm}
\includegraphics[width=1.05\columnwidth]{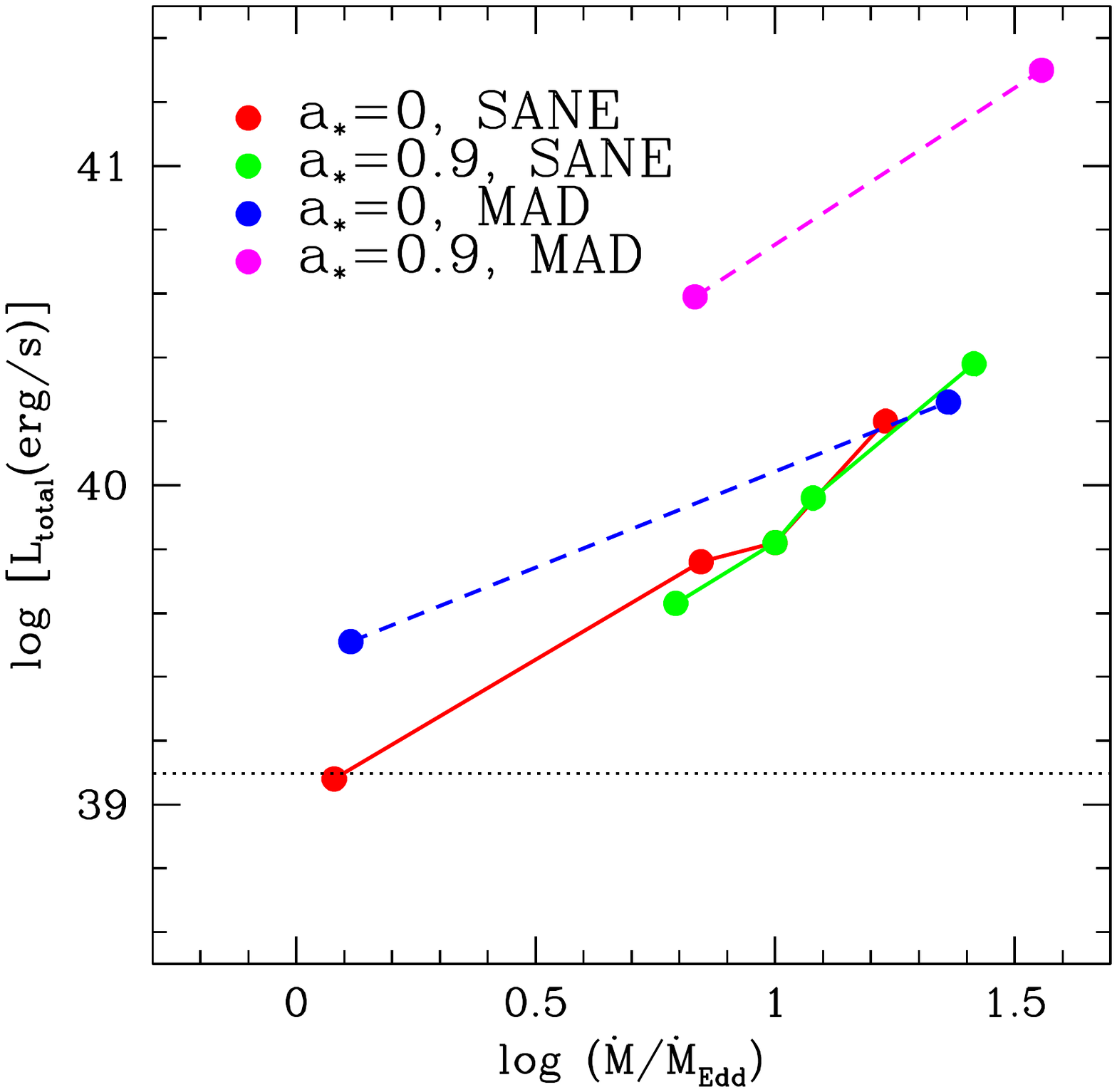}
\vspace{-2.5cm}
\caption{Total angle- and frequency-integrated radiative luminosity
  $L_{\rm rad}$ (left) and total angle-integrated net luminosity
  (radiation+jet+wind) $L_{\rm total}$\,, as a function of the mass
  accretion rate $\dot{M}$, for various sequences of SANE and MAD
  models with BH spin values of $a_*=0$ and 0.9. The horizontal dotted
  line corresponds to the Eddington luminosity.}
\label{fig:ltotal_mdot}
\end{figure*}

Figure~\ref{fig:ltotal_mdot} plots the angle-integrated radiative
luminosities $L_{\rm rad}$ and total luminosities $L_{\rm total}$
listed in Tables~\ref{tab:luma0} and \ref{tab:luma9} as a function of
the Eddington-scaled mass accretion rate. The values of $L_{\rm rad}$
are generally quite modest compared to the large apparent luminosities
shown in Figure~\ref{fig:2kev_mdot} for a favorably oriented
observer. Figure~\ref{fig:ltotal_mdot} shows a striking pattern. All
SANE models, as well as MAD models with a non-spinning BH, behave
similarly. Their radiative luminosities saturate at around $2L_{\rm
  Edd}$, which means that they are radiatively inefficient at large
accretion rates. Their total luminosities, however, scale roughly as
$(\dot{M}/\dot{M}_{\rm Edd})L_{\rm Edd}$, i.e., the efficiency
measured via the total luminosity is independent of the accretion
rate.

MAD models with a rapidly spinning BH are very different. They are
radiatively quite efficient, with $L_{\rm rad}$ scaling as $\sim
(\dot{M}/\dot{M}_{\rm Edd})L_{\rm Edd}$ even at large $\dot{M}$.
Moreover, their efficiencies are larger yet by a factor of several
when measured in terms of $L_{\rm total}$.  The distinct properties of
the MAD models with rapidly spinning BHs is almost certainly related
to the fact that they receive a powerful luminosity boost from the
spin energy of the BH. As has been shown in recent work
\citep{tchekhovskoy11,mckinney12}), energy extraction from the BH
works best when the hole spins rapidly and the magnetic field strength
approaches the MAD level.

\section{Comparison with the observed properties of ULXs}\label{sec:ULX_comparison}

\subsection{X-ray spectra}\label{ULX_spectra}

The observed X-ray spectra of ULXs show a variety of shapes, and do
not neatly fit into ``states'', unlike the spectra of sub-Eddington
stellar-mass BHs. A popular phenomenological classification of ULX
spectra includes four regimes (\citealt{sutton13},
\citealt{urquhart16}; \citealt{soria11}; \citealt{gladstone09};
\citealt{makishima07}): a) ``disk-like", well fitted by non-standard
multicolor disk models or slim disks, slightly broader than a standard
disk-blackbody; b) ``supersoft", dominated by a thermal component with
$kT < 150$\,eV; c) ``soft", dominated by a power-law with photon index
$\Gamma \sim 2-3$; d) ``hard", dominated by a power-law with $\Gamma
\sim 1-2$. The power-law component often shows a break at a
characteristic photon energy $\approx5-6$\,keV, a feature seen in both
soft and hard ULX spectra, but not in sub-Eddington stellar-mass
BHs. Also, both soft and hard ULXs often show an additional
(non-dominant) thermal component with $kT \approx 150-300$\,eV. 

When observed at sufficiently high signal to noise, many ULXs
(particularly of the supersoft and soft variety) show thermal-plasma
emission lines, absorption edges and other spectral features around 1
keV (\citealt{pinto17}; \citealt{pinto16};
\citealt{urquhart16,middleton15}), as expected for sources with strong
outflows. It is important to stress that there is a continuum of
properties between the four regimes identified above, rather than
separate classes. Transitions between different ultraluminous regimes
have been observed in a few cases
\citep{sutton13,pintore12,urquhart16}; they are more frequent than
transitions to and from quiescence.

The sequence of accretion states in the sub-Eddington regime of BH
accretion is primarily determined by changes in the accretion
rate. Here we test whether the diversity of observed spectral features
in ULXs can be explained as the result of our viewing angle, because
of various amounts of absorption and down-scattering of the hard X-ray
photons in a dense outflow. Qualitatively, we have reproduced all four
empirical regimes. For $\dot{M} \la$ few $\dot{M}_{\rm Edd}$, we find
(\S\ref{sec:mdot}) that $a_{\ast} = 0$ models with a SANE magnetic
field, seen at low inclination, produce a disk-like spectrum. For
$\dot{M} \ga$ few $\dot{M}_{\rm Edd}$, we find that inclination
effects are much more significant than changes in the accretion rate:
for $i \la 30^{\circ}$, the shape of the observed spectrum is
consistent with the hard ultraluminous regime; for $30^{\circ} \la i
\la 50^{\circ}$, with the soft ultraluminous regime; for $i \ga
50^{\circ}$, with the supersoft regime (Figures \ref{fig:fid_spectra},
\ref{fig:a0_mdot}, \ref{fig:a9_mdot}).

However, this explanation is not entirely satisfactory. Our SANE
models predict a steep decline in the observed luminosity as a
function of viewing angle, from harder to softer spectra: in
particular, hard ULXs should appear one order of magnitude brighter
than soft ULXs in the 0.3--10 keV band
(Figure~\ref{fig:fid_spectra}). This is inconsistent with
observations, which show an overlapping distribution of hard and soft
ULXs at $L_{\rm X} \approx 10^{40}$ erg\,s$^{-1}$ (Fig.~3 in
\citealt{sutton13}). In fact, there are at least two well-studied ULXs
(NGC\,1313 X-1 and NGC\,5204 X-1) that show transitions between a soft
and a hard regime, but appear brighter when softer
\citep{sutton13}. In other cases (Holmberg II X-1: \citealt{grise10};
Holmberg IX X-1: \citealt{luangtip16}), hardness changes appear
uncorrelated with luminosity changes. 

The simple fact that some ULXs show transitions between a hard and a
soft regime, or between a soft and a supersoft regime (as is the case
for example in M\,101 ULS and NGC\,247 ULS: \citealt{urquhart16}),
suggests that the viewing angle or BH spin parameter cannot be the
only parameter. A variable accretion rate may play a role, perhaps
also a variable magnetic field strength. For individual sources, disk
precession has been invoked \citep{luangtip16} to explain changes in
inclination and therefore in spectral hardness, but this explanation
is hard to reconcile with the short and irregular timescales seen for
example in M\,101 ULS \citep{soria16}.

Our MAD models predict that the apparent spectral hardness depends
both on viewing angle and, for a given angle, on $\dot{M}$, with
higher accretion rates corresponding, at least for non-rotating BHs,
to softer spectra and higher luminosities (Fig.~\ref{fig:mad}, left
panel). In this work, we have illustrated the results of MAD
simulations with the rather extreme values of $a_{\ast} = 0$ and
$a_{\ast} = 0.9$: we find that the low-spin models are a better
approximation to ULX behaviour, with a spectral turnover between 1 and
10 keV. MAD models with $a_{\ast} = 0.9$ predict too much emission
above 10 keV (regardless of accretion rate and inclination), an energy
band where observed ULX spectra drop much more steeply ({\it e.g.},
\citealt{bachetti16,bachetti13}; \citealt{rana15,walton14,walton13}).
Clearly, further work needs to be done to produce a grid of
simulations over the full range of spins and mass accretion rates, but
our first results are encouraging.

\subsection{ULX bubbles}\label{ULX_bubbles}

A powerful observational constraint we have not discussed yet is
provided by the large bubbles of ionized gas seen around several ULXs
\citep{pakull02,pakull03,pakull08,feng11}. When such ULX bubbles are
dominated by X-ray photo-ionization, the optical flux in the He II
$\lambda$4686 line provides a good proxy for the ionizing flux from
the central source \citep{pakull02}. For the photo-ionized bubble
powered by the ULX in Holmberg II, the minimum input luminosity must
be at least $L_{\rm X} \ga 4 \times 10^{39}$ erg\,s$^{-1}$ and, more
likely, $L_{\rm X} \ga 6 \times 10^{39}$ erg\,s$^{-1}$
\citep{kaaret04,pakull02}, within a factor of 2 of the apparent X-ray
luminosity of this ULX \citep{goad06,sutton13}. This rules out strong
beaming, at least for this source. It also shows that ULXs can have a
true isotropic luminosity higher than the asymptotic upper limit ($L
\approx 2.5 \times 10^{39}~{\rm erg\,s^{-1}}$) predicted by our SANE
simulations, but consistent with our MAD models at high spin
(Tables~\ref{tab:luma0} and \ref{tab:luma9}).

Other ULXs are surrounded by shock-ionized bubbles, with diameters of
$\sim$100--300 pc, powered by a collimated jet and/or fast outflows
\citep{pakull08}. The mechanical power required to inflate these
bubbles is $\sim 10^{39}$--$10^{40}$ erg\,s$^{-1}$, consistent with
the mechanical power produced in our SANE and MAD simulations
(Tables~\ref{tab:luma0} and \ref{tab:luma9}).  If the ULX photon
emission were strongly beamed, we would see many shock-ionized bubbles
without a strong central X-ray source for every ULX-associated bubble
found. This is not what is observed: most of the large shock-ionized
bubbles do contain a strong X-ray source. The number of ULX bubbles
modelled in detail is still small, but the above argument (outlined in
\citealt{pakull08}) is already a promising way to constrain the
opening angle of the polar funnel in MHD simulations.

\subsection{Optical counterparts}\label{ULX_counterparts}

Another constraint on the accretion model and its geometry comes from
the broadband emission of the optical counterpart. In X-ray binaries
and ULXs, the outer region of the accretion disk intercepts and
reprocesses a fraction of the X-ray flux from the central source,
contributing to the broad-band near-UV/optical/near-IR emission. For
sub-Eddington high-mass X-ray binaries, this contribution is usually
much lower than the emission from the massive donor star
\citep{lewin95,frank02}. In contrast, the optical emission of low-mass
X-ray binaries in outburst is dominated by the reprocessed emission of
the irradiated disk \citep{dubus99}; there is an empirical relation
\citep{vanparadijs94} between the optical luminosity of the disk, the
X-ray luminosity of the central source, and the binary period (proxy
for the disk size). 

For a standard thin disk, theoretical models ({\it e.g.},
\citealt{dubus99,king97,dejong96,vrtilek90}) and observations ({\it
  e.g.}, \citealt{russell14,soria12,gierlinski09,hynes02}) suggest
re-emission fractions of a few times $10^{-3}$.  For ULXs, the
relative contribution of disk and donor star is still an unsolved
problem
\citep{sutton14,heida14,gladstone13,grise12,tao12,tao11,copperwheat07}. In
most cases where a point-like counterpart is unequivocally identified,
its near-UV/optical/near-IR luminosity is consistent both with a
massive donor (usually a B-type supergiant) and with an outer
accretion disk (with a size of $\sim$10$^{12}$ cm) that intercepts and
re-emits $\sim$ a few $10^{37}$ erg\,s$^{-1}$ $\sim$ a few $10^{-3}$
times the apparent X-ray luminosity. There is at least one ULX, the
transient source in M\,83 \citep{soria12,long14}, where the optical
emission was proved to be from the irradiated disk, because it was
only seen when the X-ray source was bright; it requires a disk
reprocessing factor $\approx 5 \times 10^{-3}$.

\begin{figure}
\includegraphics[width=1.05\columnwidth]{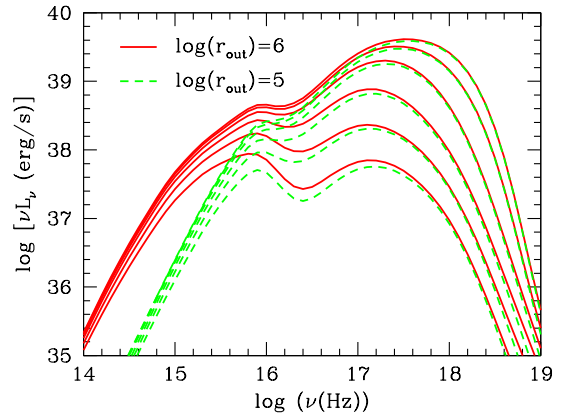}
\vspace{-3cm}
\caption{Broadband spectra of the fiducial model r010\_3d for
  observers at inclination angles (from above) of $10^\circ$,
  $20^\circ$, $30^\circ$, $40^\circ$, $50^\circ$, $60^\circ$,
  respectively. The dashed green curves correspond to the case when
  the disk is extrapolated to $r_{\rm out}=10^5$ (the default) and the
  solid red curves correspond to $r_{\rm out}=10^6$.}
\label{fig:rout}
\end{figure}

Such high levels of disk irradiation would appear to be inconsistent
with our simulated models, where the X-ray emission is strongly beamed
along the polar axis. Indeed, as the broadband spectra in
Figure~\ref{fig:rout} show, our fiducial model (dashed green curves)
produces very little emission in the optical B-band
($\log\nu\approx14.8$). Interestingly, when we postprocess the same
model with \texttt{HEROIC}, but extrapolating the disk to $r_{\rm
  out}=10^6$ rather than the default $r_{\rm out}=10^5$, the
corresponding spectra (solid red curves) have B-band luminosities
surprisingly close to the levels observed in ULXs. For a $10M_\odot$
BH, $r_{\rm out}=10^5$ corresponds to a physical outer radius of
$1.5\times10^{11}$\,cm, which is smaller than the radius $\sim
10^{12}$\,cm where optical reprocessing is believed to happen. The
model with $r_{\rm out}=10^6$ does go out to this radius, which
perhaps explains why this model agrees much better with the optical
observations.  One caveat is that the manner in which we extrapolate
the disk to large radii is fairly approximate
(\S\ref{sec:disk_extension}), so one should not take model predictions
at such radii too seriously.

%Neither our SANE nor MAD models produce enough X-ray flux at high
%inclination angles, that can be intercepted and thermalized by the
%outer disk, for plausible flaring angles. However, our simulations
%only deal with the photon emission region near the BH. We suggest that
%part of the X-ray photons emitted in the polar funnel get scattered
%more isotropically at larger distances ($\ga$10$^5 GM/c^2$), by
%optically thin gas that surrounds the ULX. This scattering should
%provide enough photons to illuminate the outer disk, and should also
%reduce the dependence of ULX luminosities on viewing angle for
%observers at infinity (as discussed in Section~\ref{ULX_spectra}).

%reflection from far above the central black hole by optically thin material ejected in a natal super-Eddington wind. Then, the higher reprocessing fractions reported for ULXs with wind-dominated X-ray spectra may be due to enhanced scattering on to the outer disc via the stronger wind in these objects. 

\section{Summary and Discussion}\label{sec:discussion}

The numerical simulations and radiative transfer calculations
presented in this paper are more detailed, and include more physics,
than previous work on ULXs. The closest comparison is the work of
\citet{kawashima09,kawashima12}, who carried out Newtonian radiation
hydrodynamics simulations and post-processed their simulated models
using a Monte Carlo code.  The present simulations are general
relativistic and include MHD, and the radiation post-processing is
more sophisticated since we solve for the gas temperature. Despite
these improvements, our results agree well with those of
\citet{kawashima12}\footnote{Note that \cite{kawashima12} use a
  different definition of $\dot{M}_{\rm Edd}$: their quoted accretion
  rates of 200$\dot{M}_{\rm Edd}$ and 1000$\dot{M}_{\rm Edd}$
  correspond to $\dot{M} \approx 11$$\dot{M}_{\rm Edd}$ and $\approx
  57$$\dot{M}_{\rm Edd}$, respectively (for $a_*=0$) in our definition
  of the Eddington accretion rate (eq.~\ref{eq:edd}).}, both in the
geometry of the flow (compare their Fig.~1) and in the computed
spectrum as a function of observer inclination (compare their
Figs.~$2-5$).  Their spectra are a little harder and slightly more
luminous than ours, but the qualitative agreement is striking.

We carried out a parameter study of ULX models as a function of the
mass accretion rate $\dot{M}$, the BH spin $a_*$, the magnetic field
strength (SANE/MAD), and the observer inclination angle $i$. For
observers at small inclination angles (pole-on view of the disk), all
the models produce super-Eddington luminosities. Even models with
$\dot{M}\approx\dot{M}_{\rm Edd}$ have X-ray ($0.3-10$\,keV)
luminosities of a few $\times10^{39}~{\rm erg\,s^{-1}}$, while models
with $\dot{M}$ equal to several $\dot{M}_{\rm Edd}$ have X-ray
luminosities above $10^{40}~{\rm erg\,s^{-1}}$ (see
Fig.~\ref{fig:2kev_mdot}). Thus, the simulaed models quite naturally
produce super-Eddington apparent luminosites for suitably oriented
observers. The large luminosities are caused by geometrical focusing,
with a slight boost from mild relativistic beaming.

While the apparent luminosities can be large, the true angle-averaged
(isotropic) radiative luminosities of the models are generally no more
than $2L_{\rm Edd}$ (see Fig.~\ref{fig:ltotal_mdot})\footnote{The
  discussion here does not include MAD models with spinning BHs, which
  are considered separately at the end of the section.}. This means
that, as $\dot{M}$ increases, the models become radiatively more and
more inefficient; for example, $\eta_{\rm rad} = 0.006$ for model
r031\_2d (Table~\ref{tab:luma0}) and $0.008$ for model r034\_2d
(Table~\ref{tab:luma9}). This result is consistent with our previous
work (e.g., \citealt{sadowski16a,sadowski_energy16}), but is in
tension with the results reported by \citet{jiang14}, who simulated a
model with $\dot{M} = 13\dot{M}_{\rm Edd}$ (converted to our
definition of the Eddington accretion rate) and found a radiative
luminosity of $10L_{\rm Edd}$, corresponding to a radiative efficiency
of 4.5\%. In contrast, our model r010\_3d, with
$\dot{M}=10\dot{M}_{\rm Edd}$, has a radiative luminosity $<2L_{\rm
  Edd}$, and an efficiency of only 0.9\%.  To compound the problem,
\citet{jiang14} find that a good fraction of their luminosity is
emitted inside 10 Schwarzschild radii, whereas in our models the
radiation is released farther out \citep{sadowski_energy16}.

The reason for the discrepancy is not clear. \citet{jiang14} used a
Newtonian code and, because they worked with cylindrical coordinates,
had a cylindrical event horizon\footnote{\citet{igumenshchev03}
  discuss the important role of the inner boundary condition for MHD
  simulations (see their Fig.~17).}. Our code is general relativistic
and models the BH horizon consistently. On the other hand,
\citet{jiang14} used a superior method to handle radiation in their
simulations, whereas our \texttt{KORAL} simulations use the simpler M1
closure scheme, although we then post-process the simulated model with
a detailed radiative transfer computation using \texttt{HEROIC}.
Interesetingly, the discrepancy between the two codes is less severe
when we consider the total luminosity:
radiation+wind+jet. \citet{jiang14} find a total luminosity of
$12L_{\rm Edd}$ and a total efficiency $\eta_{\rm total}=5.4\%$, while
we find for model r010\_3d a total luminosity of $5.3L_{\rm Edd}$ and
$\eta_{\rm total}=3.0\%$. The key difference is that our GR model
emits the bulk of its luminosity in a mechanical outflow whereas the
Newtonian model produces mostly radiation. Perhaps the vertical
advection of radiation, which \citet{jiang14} highlight in their work,
becomes less efficient with the introduction of general relativistic
dynamics in our model.

We consider next the results of our simulations in the context of
ULXs.  The range of spectra we find across our model parameter space
includes examples that resemble all the spectral states observed in
ULXs. Even the optical emission of the disk, which arises at very
large radii, appears to be roughly consistent
(Fig.~\ref{fig:rout}). In the X-ray band, one of the key observational
problems addressed in our simulations is whether the difference
between softer and harder ULX spectra is primarily due to viewing
angle or mass accretion rate.
%This is where the predictions of our SANE and MAD models for
%non-spinning (and presumably also slowly-spinning) BHs differ
%strikingly, paving the way for possible future observational tests
%between the two scenarios.

We find that the spectra of models with SANE magnetic fields are
essentially independent of Eddington ratio, and any softening of the
spectrum is purely a result of an increasing viewing angle
(Figs.~\ref{fig:a0_mdot}, \ref{fig:a9_mdot}).  MAD models around
non-spinning BHs, by contrast, predict a dramatic spectral softening
with increasing accretion rate, even for face-on observers
(Fig.~\ref{fig:mad}, left panel). This is caused by the optical depth
of the polar outflow increasing and a scattering photosphere
developing inside the funnel.  A qualitatively similar softening of
the observed spectrum for increasing accretion rates was also found by
\cite{kawashima12}, for similar reasons (more severe down-scattering
in a denser wind); quantitatively, the softening effect is more
pronounced in our zero-spin MAD models.

As a consequence of the above effect, our MAD models predict that the
apparent luminosity distribution of soft ULXs should largely overlap
that of hard ULXs, in agreement with observations
\citep{sutton13}). SANE models, on the other hand, predict that softer
ULXs should always appear systematically fainter.  The emergence of a
photosphere in the polar funnel at very high accretion rates in the
MAD models supports the suggestion of \cite{soria16} (based on simple
analytic approximations) that ultraluminous supersoft spectra may be
caused by extremely super-Eddington accretion rates, even for
low-inclination viewing angles.

There is, however, one serious problem when applying our models to ULX
observations. All the simulated models have geometrically thick disks
with narrow funnels, requiring the observer to be located within
$20-30^\circ$ of the poles to see the bright hard emission from gas
near the BH. Off-axis observers see softer spectra with luminosities
that rapidly fall below the defining luminosity limit of a ULX. The
strong geometrical beaming implies that the observed ULXs should
represent only $\sim10\%$ of a larger population, the remaining
$\sim90\%$ being beamed away from us. The question then is: why have
we not seen the ULX bubbles associated with these latter off-axis
objects?  The radiation from the bubbles should not be beamed and
therefore should be visible, independent of orientation. The absence
of a large population of ``orphan bubbles'' strongly suggests that the
geometrical beaming in our simulated ULX models is too large.

There is no obvious solution to the above discrepancy. The narrow
funnels in our simulations are caused by a strong radiatively-driven
wind which originates close to the BH. This wind restricts the range
of angles over which the hot gas near the BH is visible to a distant
observer. Even models with $\dot{M}\sim 1\dot{M}_{\rm Edd}$ (e.g.,
model r012\_3d) show pronounced beaming, as does Fig.~1 in
\citet{kawashima12}. The beaming is stronger, and shows less $\dot{M}$
dependence, than the empirical model of \cite{king09}. The opening
angle of the funnel is determined by the shape of the thick accretion
disk at small radii. It is possible that the initial torus with which
we initialize the simulations causes the disk to be too thick, and the
funnel to be too narrow. It would be worthwhile to investigate how the
initial conditions of the simulations affect disk thickness and degree
of beaming.

A general result from this work, which should apply to all
super-Eddington systems, not just ULXs, is that the angle-integrated
radiative luminosity is capped at a few $L_{\rm Edd}$, even when
$\dot{M}\gg\dot{M}_{\rm Edd}$, whereas the total
radiative-plus-mechanical luminosity is much larger, $L_{\rm tot}\sim
(\dot{M}/\dot{M}_{\rm Edd})L_{\rm Edd}$ (see also
\citealt{sadowski_energy16}). Mechanical feedback should thus be very
strong in super-Eddington systems. Does this feedback prevent the
occurrence of super-Eddington AGN altogether? Does it prevent BH seeds
from growing at super-Eddington rates in the early universe?  These
are open and interesting questions for future research.

We turn finally to the two models we simulated of super-Eddington MAD
accretion on rapidly spinning BHs, viz., models r014\_3d and r015\_3d.
These two models behave very differently from the other models we have
discussed so far, and this regime of accretion has unique properties,
as noted already by \citet{mckinney14,mckinney15}.  The radiative
luminosity is much higher, and the accretion is radiatively efficient
even at large $\dot{M}$ (Table~\ref{tab:luma9} and
Fig.~\ref{fig:ltotal_mdot}, right panel). The spectrum is very hard
and extends well above 100\,keV (Fig.~\ref{fig:mad}, right panel). The
total luminosity, including the mechanical energy carried out in an
outflow, is several times larger than the already large radiative
luminosity, giving total luminosity efficiencies $\sim70\%$
(Table~\ref{tab:luma9}, compare with \citealt{tchekhovskoy11}, who
obtained $>100\%$ efficiency for a non-radiative MAD model). All of
these unusual properties can be traced to the fact that these systems
are able to use the MAD-level magnetic field to tap the spin energy of
the BH, thereby producing powerful relativistic jets and strong
beaming effects.

The spectra of the two large-BH-spin MAD models do not resemble the
spectrum of any ULX. This suggests that ULXs either do not reach the
MAD state or do not have rapidly spinning BHs. The former possibility
is unattractive since we argued earlier that slowly spinning BHs with
MAD accretion do fit ULX observations; specifically, they explain
luminous systems with soft spectra. Is it possible that BHs in ULXs do
not have large spin values? Could the large mass accretion rate that
is characteristic of the super-Eddington regime cause a rapid spin
down of the holes?

Even though spinning MAD models do not appear to descibe ULXs, the
features they show are very promising for modeling TDE systems such as
Swift J1644+57, which \citet{tchekhovskoy14} argue was produced by a
spinning BH with a MAD-level magnetic field. This regime of accretion
also appears promising for understanding the high energy spectra of
FSRQ blazars \citep{maraschi03}.

\section{Acknowledgements}

The authors thank Magdalena Menz for help with the CHIANTI opacities
used in the present work. RN was supported in part by NSF grant
AST1312651, NASA grant TCAN NNX14AB47G, and the Black Hole Initiative
at Harvard University, which is supported by a grant from the John
Templeton Foundation.  AS acknowledges support by NASA through
Einstein Postdoctoral Fellowship number PF4-150126 awarded by the
Chandra X-ray Center, which is operated by the Smithsonian
Astrophysical Observatory for NASA under contract NAS8-03060. The
authors acknowledge computational support from NSF via XSEDE resources
(grant TG-AST080026N), and from NASA via the High-End Computing (HEC)
Program through the NASA Advanced Supercomputing (NAS) Division at
Ames Research Center. RS acknowledges hospitality at the National
Astronomical Observatory of China (University of the Chinese Academy
of Sciences), in Beijing, during part of this work.

\bibliography{ms.bib}

\end{document}